\documentclass[twocolumn]{aastex631}

\usepackage{amsmath}
\usepackage{amssymb}
\usepackage{subfigure}
\usepackage{natbib}
\usepackage{hyperref}
\usepackage{breakurl}
\usepackage{xcolor}
\usepackage{enumitem}
\setlist{leftmargin=5.5mm}

\newcommand\src{PSR J1048+2339}
\newcommand{\mail}{lilirayhk@phys.ncku.edu.tw}

\newcommand{\lum}{\,erg\,s$^{-1}$}
\newcommand{\kms}{\,km\,s$^{-1}$}
\newcommand{\cm}{\,cm$^{-2}$}

\newcommand{\bb}[1]{\textcolor{black}{#1}}
\shorttitle{}
\shortauthors{Li et al.}

\begin{document}
\title{Sharpening the Mass Measurement of \src\ via High-Cadence Photometry}

\correspondingauthor{Kwan-Lok Li}
\email{\mail}

\author[0000-0001-8229-2024]{Kwan-Lok Li}
\affiliation{Department of Physics, National Cheng Kung University, 70101 Tainan, Taiwan}

\author{Ka-Yui Au}
\affiliation{Department of Physics, National Cheng Kung University, 70101 Tainan, Taiwan}

\author{Lupin C. C. Lin}
\affiliation{Department of Physics, National Cheng Kung University, 70101 Tainan, Taiwan}

\author{C. Y. Hui}
\affiliation{Department of Astronomy and Space Science, Chungnam National University, Daejeon, Republic of Korea}

\author{Albert K. H. Kong}
\affiliation{Institute of Astronomy and Institute of Space Engineering, National Tsing Hua University, Hsinchu 30013, Taiwan}
\affiliation{Institute of Space Engineering, National Tsing Hua University, Hsinchu 30013, Taiwan}

\author{Jumpei Takata}
\affiliation{Department of Astronomy, School of Physics, Huazhong University of Science and Technology, Wuhan 430074, People’s Republic of China}

\begin{abstract}
Neutron stars are among the densest objects in the universe, and the mass measurements are crucial to the understanding of their equation of state. Pulsars in tight binaries can be weighed dynamically, but accurate mass estimates are often hampered by uncertainties in orbital inclinations. In this paper, we report on multi-epoch optical monitoring of the pulsar binary, \src, using the 0.5-m RIFT telescope at \textit{Lulin} Observatory. Our observations reveal frequent optical flares with eclipses caused by the companion. Using the eclipse duration with archival photometric data and orbital ephemeris in the literature, we constrain the inclination to $78\fdg0\pm{1\fdg0}$ and the pulsar mass to $1.79^{+0.09}_{-0.08}$ solar masses. This measurement is among the most precise for any redback system to date, demonstrating a novel way to improve the pulsar mass measures with small aperture optical telescopes. Furthermore, the optical eclipse duration is significantly shorter than that measured in M/GeV $\gamma$-rays, possibly suggesting that a fraction of the eclipsed $\gamma$-ray emission originates from the apex of the intrabinary shock. Based on these constraints, we estimate the magnetic field strength of the companion to be 30--170~G. This is consistent with the high field strength implied by the X-ray orbital modulation observed. 
\end{abstract}

\section{Introduction}

In compact millisecond pulsar binaries, the low-mass companion stars are being ablated by strong radiations due to their proximity to the neutron stars. These systems are named ``black widow'' (with semi-degenerate companions) or ``redback'' (with non-degenerate companions) because of the similarity to the mating behavior of the spider species \citep{2013ApJ...775...27C,2019Galax...7...93H}. The companions in these binaries are tidally locked so that the same sides always face to the radiating neutron stars and get irradiated to higher temperatures. In addition, the shapes of the companions can be ellipsoid-like due to the strong gravity of the neutron stars causing ellipsoidal variation. These result in orbital modulations in the optical bands, which can be used to extract information of the orbital parameters of the binaries, and hence, the masses of the neutron stars \citep{2012ApJ...760L..36R,2016ApJ...833..143L,2019ApJ...872...42S}.

\src\ is a 4.66-ms redbacks millisecond pulsar in a tight binary of an orbital period of 6.01 hours, discovered in radio by Arecibo and Green Bank Telescope (GBT; \citealt{2016ApJ...819...34C,2016ApJ...823..105D} and later in M/GeV $\gamma$-rays by the \textit{Fermi} Large Area Telescope (LAT; \citealt{2013ApJS..208...17A}. It is also one of a few redback systems that exhibits variable irradiation \citep{2018ApJ...866...71C}, of which the power can increase up to a factor of 6 in just two weeks \citep{2019A&A...621L...9Y}. 

There were several mass measurement attempts for \src\ using optical observations together with the ephemeris of the pulsar \citep{2019A&A...621L...9Y,2019ApJ...872...42S,2021A&A...649A.120M,2025MNRAS.536.2169S}. Despite the excellent quality of the data, a wide possible mass range of $M_\mathrm{NS}=1.08$--$2.3M_\sun$ was inferred because of the unknown orbital inclination angle. A recent breakthrough came from the \textit{Fermi}-LAT discoveries of the $\gamma$-ray eclipses of two black widows and five redbacks, including \src. If the eclipses are caused by occultation by the companion, the mass of \src\ can be further constrained to $M_\mathrm{NS}=1.44$--$1.72M_\sun$ \citep{2023NatAs...7..451C}.

In this paper, we report on the high-cadence optical photometric data of \src\ taken by the \textit{Robotic Imager For Transients} (RIFT). Unusual ``flare eclipses'' were observed in the data, with which we further constrained the pulsar mass and measured the magnetic field strength of the companion by comparing the eclipse durations detected by RIFT and \textit{Fermi}-LAT.

\section{RIFT Observations}

RIFT is a 0.5-m telescope at the \textit{Lulin} Observatory in Taiwan. It is a corrected Dall-Kirkham (CDK) telescope manufactured by the PlaneWave Instruments with a focal ratio of $f/4.5$. The telescope is mounted on the L-series equatorial mount designed by the same company with a tracking accuracy of $<0\farcs3$ in 5 minutes. Together with the Andor Technology camera, iKon-L~936, with a $2048\times2048$ charge-coupled device (CCD) array (or $27.6\,{\rm mm}\times27.6\,{\rm mm}$), the $f/4.5$ CDK telescope has a field of view (FoV) of $41\farcm7\times41\farcm7$, equivalent to 0.44~deg$^2$, with a pixel scale is 1\farcs22 per pixel. The robotic telescope system is operated by a \texttt{Python} script developed by our own team with the Astronomy Common Object Model (\texttt{ASCOM}) interface.

We carried out a three-month observing campaign for \src\ using the 0.5-m telescope in 2023. From 1st February to 28th April 2023, about 4,000 unfiltered images of \src\ were taken in 48 nights of good weather and grey/dark sky. Each image has an exposure of 200 seconds, equivalent to 0.9\% of the orbital period. 

\subsection{Data Reduction}
The Image Reduction and Analysis Facility (\texttt{IRAF}) was used to reduce the RIFT observations with standard reduction processes, including bias-frame subtraction, dark-frame subtraction and flat-frame correction. The astrometry was then performed by \texttt{astrometry.net} \citep{2010AJ....139.1782L} to obtain a reliable world coordinate system (WCS) solution for each image. As the field of the \src\ is not crowded, aperture photometry was applied using \texttt{phot} with an aperture radius of 3 pixels (or 3\farcs7) and an annulus background region (inner radius: 15 pixels; outer radius: 25 pixels). We further employed differential photometry to eliminate the effect from the varying weather condition. A bright ($R_1=14.64$~mag) and nearby (less than 1\arcmin\ away from \src) field star, USNO-B1.0 1136$-$0178206 \citep{2003AJ....125..984M} was chosen as the comparison star. Comparing with a check star in the field, the differential magnitude light curve of the comparison star is very stable with a fluctuation less than 0.02~mag. The uncertainties of the differential magnitude light curve of \src\ were estimated by, $\sigma=\sqrt{\sigma_{i,t}^2+\sigma_{i,c}^2}$, where $\sigma_{i,t}$ and $\sigma_{i,c}$ are the uncertainties of the instrumental magnitudes of the target and the comparison star, respectively. As the weather condition of \textit{Lulin} could be unstable throughout a night, we further selected the data by removing those with uncertainties larger than 0.2~mag, equivalent to a 5$\sigma$ signal-to-noise ratio. 

\subsection{Flare Eclipses}

\begin{figure*}
\includegraphics[width=0.9\textwidth]{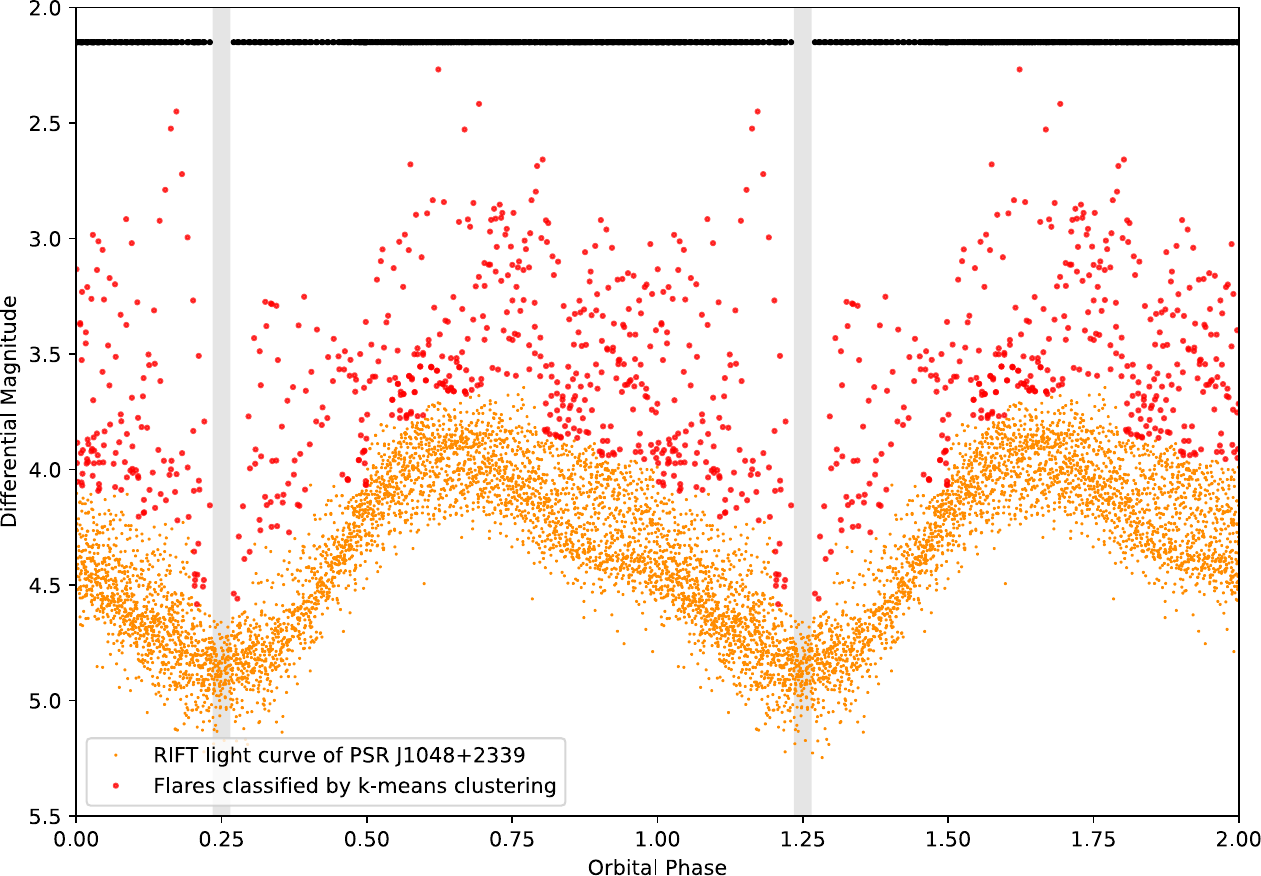}
\centering
\caption{The RIFT unfiltered optical light curve of \src. The light curve was phased on the orbital period of $P_\mathrm{orb}=0.250519045$ days. The y-axis is the unfiltered differential magnitude, which can be roughly transfer into apparent magnitude by adding 14.64~mag. The smaller orange data points shows the quiescent state, while the bigger red data are those classified as flaring data by k-means clustering. The upper horizontal black lines are the projections of the flaring observations, which clearly show eclipses at phase 0.25 or 1.25. This is in line with the inferred eclipse duration of 0.0325 orbits (or $11\fdg7$), indicated by the two vertical grey shadows. Two identical cycles are shown for clarity.}
\label{fig:j1048_lc}
\end{figure*}

\begin{figure}
\includegraphics[width=0.5\textwidth]{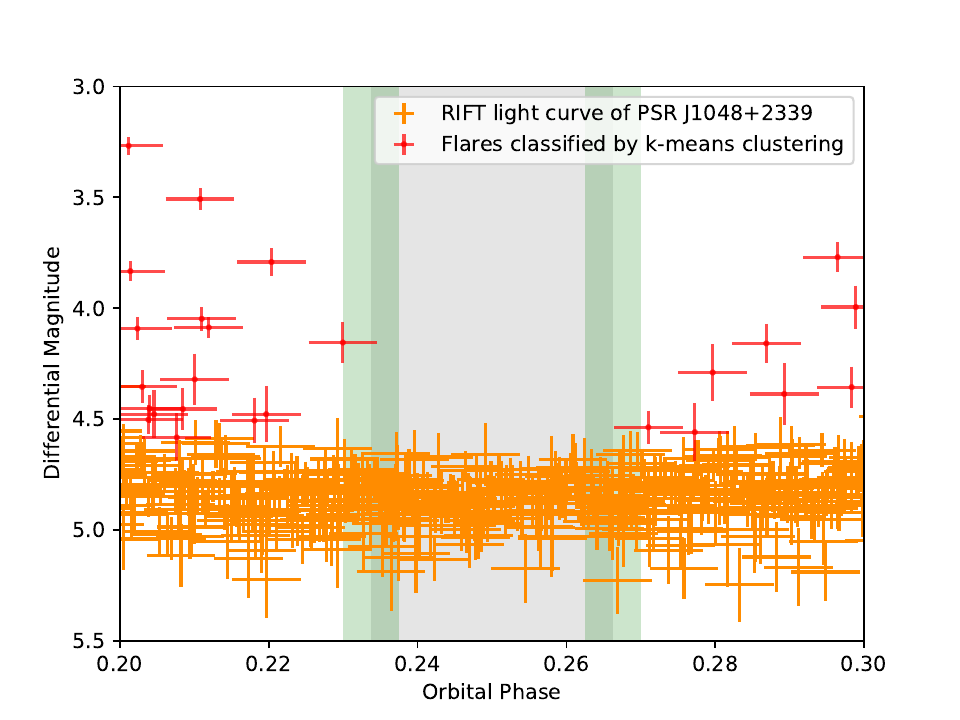}
\centering
\caption{\bb{A zoomed-in version of Figure \ref{fig:j1048_lc} around the eclipse region. The transparent green shadow indicates the estimated range of the eclipse duration. The errors of both axes are also shown here for a better comparison, i.e., 1$\sigma$  errors of $\sim0.1$~mag for the magnitudes and about 200 seconds or 0.01 phases for the orbital phases.}}
\label{fig:j1048_elcipse_lc}
\end{figure}

Figure \ref{fig:j1048_lc} shows the stacked RIFT light curve of \src\ folded on the orbital period of 0.250519045 days \citep{2016ApJ...823..105D} with phase zero defined as the epoch of the pulsar ascending node (i.e., the pulsar is going away in the view from Earth at the highest radial speed; \bb{individual RIFT light curves are shown in Figure \ref{fig:individual_obs}}). A single peak around the epoch of the superior conjunction of the pulsar (i.e., the companion is behind the neutron star), a signature of the irradiation effect, was observed. The amplitude of the orbital modulation was varying, indicating that the heating was changing throughout the observing campaign, though it was not as dramatic as seen in the previous studies \citep{2018ApJ...866...71C,2019A&A...621L...9Y}. 

In addition to the orbital modulation, there were optical flares detected once every few days (i.e. about 10\% of the observations show flares; \bb{see the online figure set associated with Appendix \ref{app:rift}}). The flare amplitudes range from less than 0.5 mag to larger than 2.0 mag with durations on time-scales from minutes to hours. Most interestingly, the flares appeared at every orbital phase, but a narrow interval at the inferior conjunction, reminiscent of pulsar eclipses (flare eclipses hereafter). 

We adopted the k-means clustering method on the stacked light curve to perform classification statistically. The stacked phased light curve was first divided into 10 parts so that each of which contains only simple orbital variation (compared to a sinusoidal modulation). Each segment was then classified into two groups, flaring or quiescence, based on their differential magnitudes using the \texttt{kmeans2} algorithm of the \texttt{SciPy} package in \texttt{Python} (Figure \ref{fig:j1048_lc}). A total number of 520 flaring data were found out of the 4377 observations. While we notice that there could be some possible flaring observations missed in the phase interval of 0.3--0.5, they are all marginal cases and we conclude that it does not affect the main result much. No flare (including those undetected by k-means but marginally seen; Figure \ref{fig:j1048_lc}) was found at phase 0.25 with an interval of around 0.03 orbits. We also checked the light curves individually and at least six eclipse ingresses or egresses were clearly observed on 2023 February 16, 21, 22, 28, April 9, and 14 \bb{(Figures \ref{fig:individual_obs}i, n, o, u, ak, and ao}, respectively.

\subsubsection{Eclipse Duration}
The upper limit was measured with the flare eclipse ingress observed on 2023 April 14, which is one of the most prominent flares observed by RIFT. There were three data points taken around the eclipse that night, of which the differential magnitudes are $m=4.16\pm0.09$, $4.68\pm0.13$, and $4.66\pm0.14$~mag at phases $\phi=0.2300$, $0.2396$, and $0.2492$, respectively (\bb{with separations of about 200 seconds}; \bb{Figure \ref{fig:individual_obs}ao}). By comparing the data with the stacked light curve (the grey curve in the figure), it is clear that the first and brighter data point was taken when the pulsar is just outside the eclipse region. This data is also the latest eclipse ingress data observed throughout the RIFT dataset. However, the eclipse could have started during this last flare observation, and thus we conservatively set the upper limit as $\theta<0.040$ orbits (or 14.4\arcdeg) using the the phase of the data (i.e., $\phi=0.2300$). 

The above method is straightforward but it is not applicable to the lower-limit measurement because the real eclipse duration can always be shorter than the observed eclipse duration (i.e., we do not know whether an eclipsed data is really caused by eclipse or just the variability of the system). A statistical approach was therefore used: 

\begin{enumerate}[label=(\alph*)]
\item We visually examined the RIFT light curves and recorded the duration of every flare (i.e., the duration distribution of the flares). While k-means clustering has classified the light curves and identified many short-lived (e.g., single-point) weak flares, we aimed to find flares that are long-lived and prominent through this visual check for conservative purpose (i.e., marginal detections were ignored). 

\item We randomly placed these known flares in the time windows of the observations \bb{(with time resolutions of roughly 200 seconds)} to have a simulated dataset. 
\item Assuming that the system has an eclipse duration of $D_e$, we removed all the flares that were placed within the assumed eclipse region (a symmetric eclipse was assumed). 
\item We repeated steps (b) and (c) 100,000 times, and measure the fraction (chance probability) of the simulated data that reproduced the observed RIFT eclipses. If the chance probability is low, then the corresponding $D_e$ is unlikely the true eclipse duration (i.e., $\theta$). 
\end{enumerate}

Note that the chance probability is in fact a cumulated probability (i.e., $P[\theta\le D_e]$) here. For example, a simulated data that reproduces the RIFT-observed eclipses with $D_e=0.01$ orbits can replicate those with $D_e=0.02$ orbits, but not vice versa. Therefore, $P$ must be accumulating with $D_e$. We defined the lower limit of the eclipse duration ($D_{e,low}$) as $P[\theta\le D_{e,low}]=0.1$, meaning that there is just a 10\% chance that the RIFT result can be reproduced outside the defined limit. 

\begin{figure}
\includegraphics[width=0.45\textwidth]{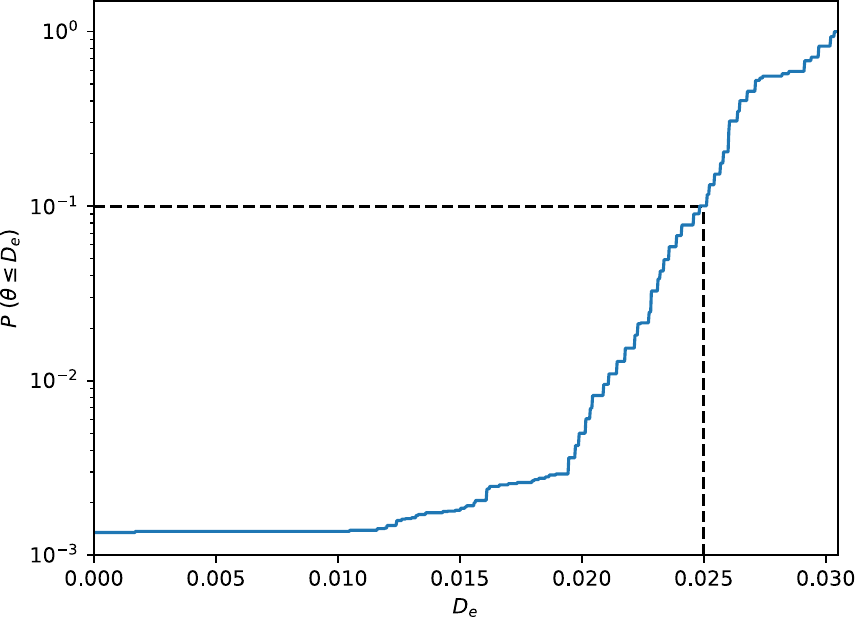}
\centering
\caption{The cumulative distribution function for the chance probability of an eclipse duration $\bm{D_e}$ reproducing the RIFT-observed eclipses. The dashed lines indicates the lower limit of the eclipse duration of \src, below which there is only a 10\% chance of reproducing the observed result.}
\label{fig:lower_limit}
\end{figure}

In total, 39 flares were found with durations ranging from 4 to 58 consecutive data points (each point presents a 200-second exposure; \bb{see Figures \ref{fig:individual_obs}an and z}, respecitvely). 
Figure \ref{fig:lower_limit} shows the result of (d) from $D_e=0$ (no eclipse) to 0.03 orbits (observed duration) with 1,000 steps. The simulations give $D_{e,low}=0.025$ orbits (or 9.0\arcdeg), making the range of the eclipse duration 0.025--0.040 orbits (or 9.0--14.4\arcdeg). 

\subsubsection{Detection Significance of the Flare Eclipse}

The lower-limit estimation method aforementioned can also be used to compute the detection significance of the flare eclipse with $P[D_e=0]$. Out of 100,000 generated datasets, 135 reproduced the eclipses observed by RIFT (no flare detection in $\theta\le0.03$ at phase 0.25), indicating a chance probability of 0.135\%, equivalent to a significance of 3.2$\sigma$. However, these 135 simulated eclipses are not necessarily symmetric by definition. If symmetry is considered, the number will be reduced to 23, equivalent to a 3.7-$\sigma$ detection.

\section{Eclipsing Light Curve (ELC) Modelling}

\begin{table*}
\centering 
\caption{The model and derived parameters of \src}
\begin{tabular}{@{}lccc}
\hline
& Unit & \multicolumn{2}{c}{Median value with the 68\% credibility interval}\\
\hline
Model parameters & & \bb{Eclipse duration constrained} & \bb{Eclipse duration unconstrained}\\
\hline
Mass ratio ($Q$) & - & $4.86\pm0.13$ & \bb{$4.86\pm0.12$} \\
Binary inclination ($i$) & [degree] & $78.0\pm1.0$ & \bb{$77.9^{+11.6}_{-11.4}$}\\
Roche-lobe filling factor ($f$) & - & $0.90\pm{0.02}$ & \bb{$0.91^{+0.06}_{-0.05}$} \\
Irradiation power ($L_\mathrm{irr}$) & [$\log$(\lum)] & $33.27\pm{0.09}$ & \bb{$33.28\pm{0.10}$}\\
Companion temperature ($T_\mathrm{eff}$) & [K] & $4140\pm150$ & \bb{$4140\pm140$}\\
\hline
Derived parameters\\
\hline
Pulsar mass ($M_\mathrm{NS}$) & [$M_\odot$] & $1.79^{+0.09}_{-0.08}$ & \bb{$1.81^{+0.26}_{-0.15}$}\\
Companion mass ($M_\mathrm{C}$) & [$M_\odot$] & $0.37\pm0.01$ & \bb{$0.37^{+0.05}_{-0.03}$}\\
Stellar separation ($a$) & [$R_\odot$] & $2.16\pm0.03$ & \bb{$2.17^{+0.10}_{-0.06}$}\\
\hline
\label{tab:elc}
\end{tabular}
\end{table*}

The orbital inclination ($i$), the component separation ($a$), and the size of the companion star of \src\ are tightly correlated by the eclipse duration measure. Provided that the mass ratio ($Q$; the neutron star mass to companion mass ratio) is known, the component separation projected onto the line of sight can be obtained through pulsar timing \citep{2016ApJ...823..105D,2021A&A...649A.120M}. This means that the separation is a function of the inclination. For compact binary systems like \src, the degrees of the stellar distortion increase with sizes according to the Roche equipotential surfaces. By modelling the optical light curves of \src, we can investigate the stellar distortion, and hence, the Roche equipotential surface. Through the known Roche equipotential surface, the companion size and the component separation are also correlated, making the former parameter a function of the inclination as well. Therefore, measuring the eclipse duration provides a strong constraint on the orbital inclination, and hence, the neutron star mass. 

Using the Eclipsing Light Curve (ELC) code \citep{2000A&A...364..265O}, we re-modeled the $g^\prime$- and $r^\prime$-band photometric data of the \textit{Lulin} One-meter Telescope (LOT) obtained by \cite{2019A&A...621L...9Y} with the RIFT eclipse duration (i.e., $\theta=0.0325\pm0.0075$ orbits) to measure the pulsar mass. In \cite{2019A&A...621L...9Y}, there were two groups of observations, in which the optical variability was dominated by the ellipsoidal variation and the pulsar heating, respectively. We adopted the dataset dominated by the ellipsoidal variation that provides more information on the stellar distortion. \bb{We note that the LOT light curves are slightly asymmetric, as mentioned in \cite{2019A&A...621L...9Y}. The feature could be modelled by adding hot/cold spots on the stellar surface, but the fitting result will then be more model-dependent as well. For a conservative result, we decided not to assume any hot/cold spots on the surface. This will cause some deviations between the model and the data, and subsequently increases some parameter uncertainties. However, the result will be more defensive.}

In the ELC models, the stellar distortion is quantified by the Roche-lobe filling factor ($f$) that is defined as the ratio between the radius of the companion in the pulsar direction and the distance from the companion to the inner Lagrangian point (L$_1$). The projected semi-major axis of the pulsar’s orbit was set to $a\sin i/(1+Q)=0.836122(3)$ light-seconds based on the result of the Arecibo and GBT radio timing observations \citep{2016ApJ...823..105D}. We also employed the radial velocity semi-amplitude of the companion obtained by the \textit{Very Large Telescope} (VLT; \citealt{2021A&A...649A.120M}), which is $K_2=353.5\pm6.5$\kms\ with the $K$-correction \citep{1988ApJ...324..411W} applied (see Appendix \ref{app:kcorr}). 
On the basis of the Very Large Telescope (VLT) spectroscopic observations \citep{2021A&A...649A.120M}, we took $T_\mathrm{eff}=4000\pm100$\,K (or K8V type) as the initial guess for the surface temperature of the companion (and $T_\mathrm{eff}$ will still be one of the fitting parameters in the model). Finally, an optimizer based on the Differential Evaluation Markov Chain Monte Carlo (DE-MC) approach \citep{2006S&C....16..239T} was applied for parameter estimation. 

\begin{figure}
\includegraphics[width=0.45\textwidth]{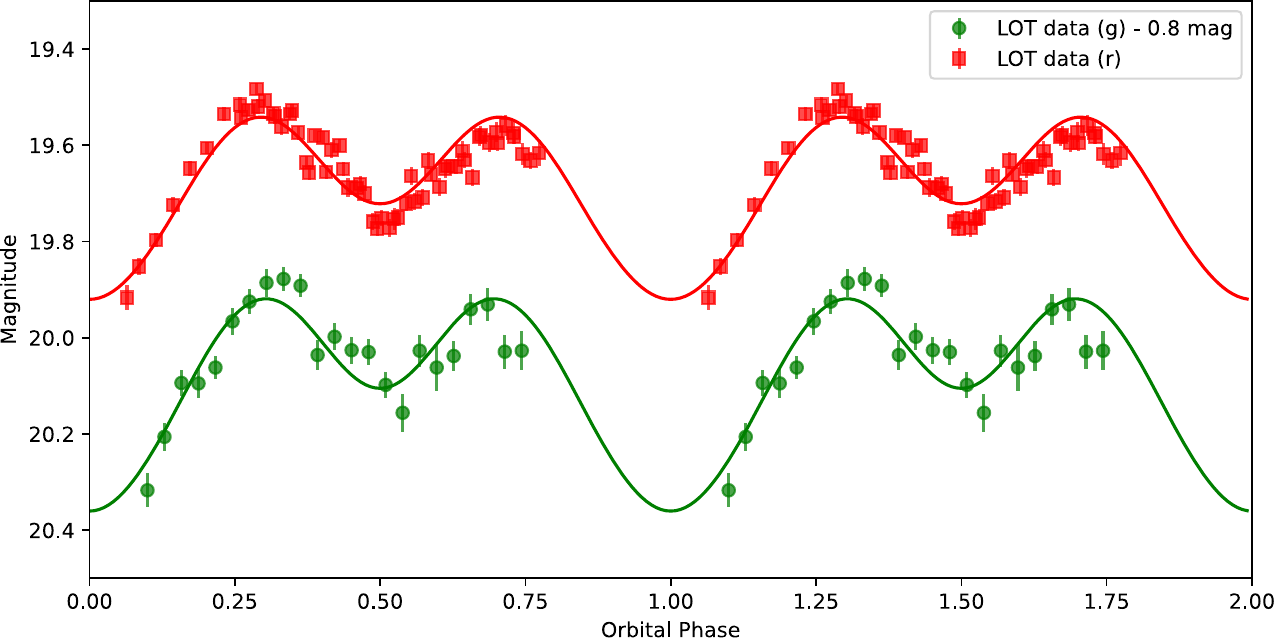}
\centering
\caption{The \textit{Lulin} LOT $\bm{g^\prime}$- and $\bm{r^\prime}$-band light curves of \src\ with the corresponding ELC models. The input model parameters can be found in Table \ref{tab:elc}. Two identical cycles are shown for clarity.}
\label{fig:elc_lc}
\end{figure}

\begin{figure*}
\includegraphics[width=\textwidth]{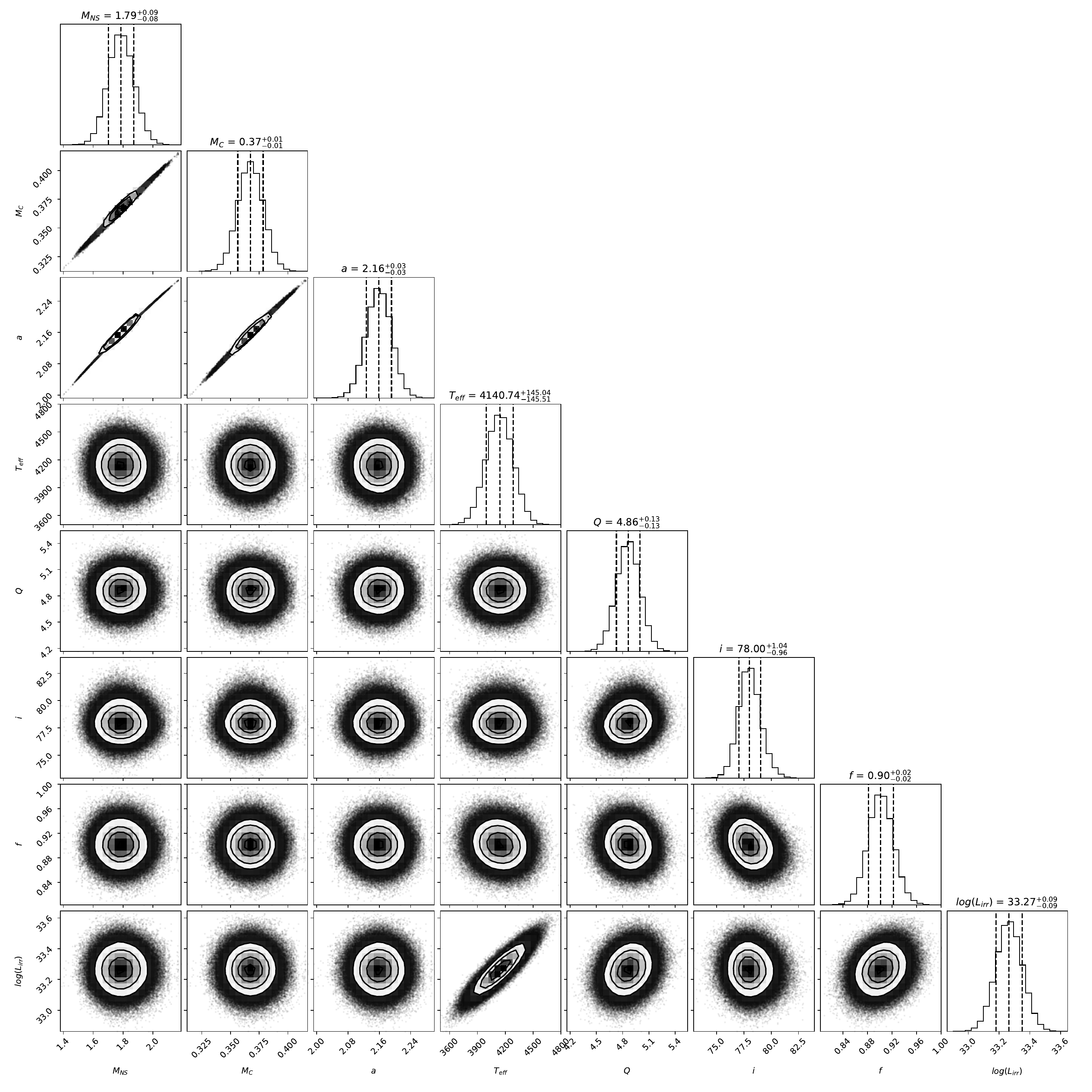}
\centering
\caption{The one/two-dimension projected posterior probability distributions of the ELC model/derived parameters of \src. In the two-dimensional projections, the 1$\sigma$, 1.5$\sigma$, and 2$\sigma$ contours are shown. In the one-dimensional plots, the quantities of the median values as well as the 68\% credibility intervals are displayed on the tops and also indicated by three vertical dashed lines in the figures, respectively.}
\label{fig:corner2}
\end{figure*}

There are five fitting parameters in the ELC model, including the mass ratio ($Q$), binary inclination ($i$), Roche-lobe filling factor ($f$), irradiation power ($L_\mathrm{irr}$ with an assumption of a point-like irradiation source), and surface temperature ($T_\mathrm{eff}$). The \texttt{demcmcELC} optimizer in the ELC code was used for parameter estimation with 30 chains evolving for 50,000 generations in the parameter space that is sufficiently large to contain all the physically possible configurations (Figure \ref{fig:corner2}). By discarding the first 33,000 generations as burn-in samples, we used the \texttt{Python} package, \texttt{corner}, to quantify the results and show the projected posterior probability distributions of the model/derived parameters (Figure \ref{fig:corner2}). Table \ref{tab:elc} shows the model and derived parameters of \src\ with 68\% credibility intervals, and Figure \ref{fig:elc_lc} shows the LOT data with the ELC model light curves. All binary parameters are well fitted, including the inclination and the Roche-lobe filling factor, i.e., $i=78\fdg0\pm{1\fdg0}$ and $f=0.90\pm0.02$, resulting in a precise neutron star mass measure of $M_\mathrm{NS}=1.79^{+0.09}_{-0.08}M_\sun$. This is among the most accurately measured pulsar masses in redback systems \citep{2018ApJ...859...54L,2019ApJ...872...42S,2023NatAs...7..451C,2024MNRAS.528.4337D,2026MNRAS.545f2173P}. 
\bb{We also had another similar ELC run with no eclipse duration constraint to study the impact of the flare eclipse in the fit. The best-fit parameters of both models are all consistent with each other. With the eclipse duration constraint, the overall precision is significantly higher (e.g., the errors of the pulsar mass are at least two times lower), although the uncertainties are slightly larger in $Q$ and $T_\mathrm{eff}$ likely due to the randomness of the sampling process.}

\section{Discussion and Conclusions}

\subsection{Flare Eclipse Interpretation}

\begin{figure*}
\includegraphics[width=0.9\textwidth]{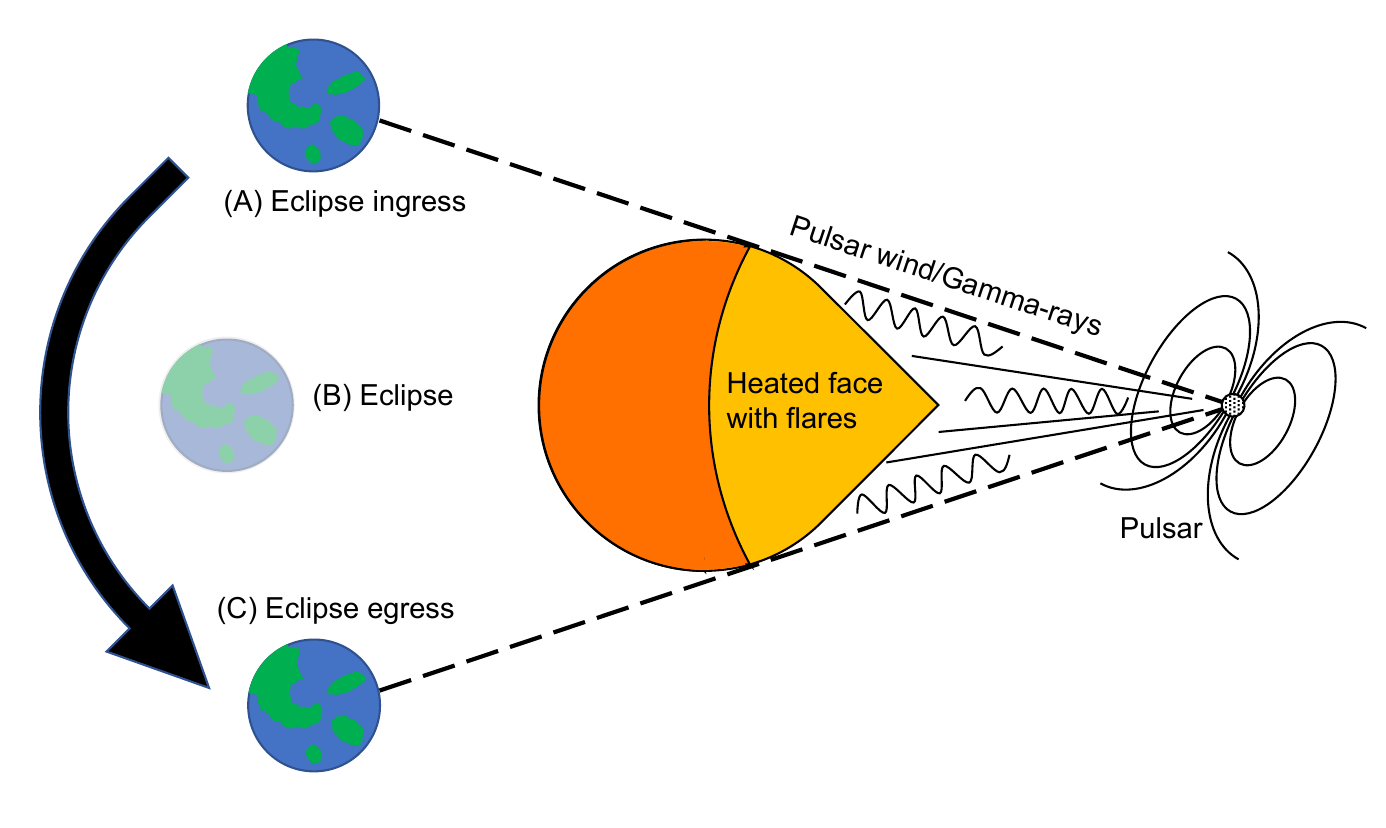}
\centering
\caption{A symbolic diagram for the origin of the flare eclipses. In the model, the flares originate from the surface of the companion heated by the pulsar wind and the gamma-ray radiation. (A) It is the epoch of the eclipse ingress when the pulsar, the surface of the companion and Earth form a straight line. (B) After the eclipse ingress, only the ``night'' side can be seen from Earth. (C) The ``day'' side (or the heated face) becomes visible to Earth again after the eclipse egress. An edge-on system is presented in the figure for simplicity, but it is not a necessary condition for the flare eclipses. }
\label{fig:j1048_sym}
\end{figure*}

Flares of \src\ have been previously observed in both X-rays and optical bands \citep{2019A&A...621L...9Y,2018ApJ...866...71C}. However, it remains unclear whether the X-ray and optical flares are related as no simultaneous pairs have been observed yet. Since the optical flares concentrated on the heated side of the companion star, the origin was proposed to be an enhanced heating effect related to the magnetic activity of the companion. This is in line with the RIFT observations that show no flares in eclipses during which the heated side was invisible to Earth, meaning that the observed flare eclipses have the same duration as that of the pulsar eclipse (Figure \ref{fig:j1048_sym}).

An alternative picture is that the optical flares were from the intrabinary shock between the pulsar and the companion star \citep{2014AN....335..313R,2014ApJ...785..131T,2014ApJ...797..111L}. 
\bb{In \cite{2021A&A...649A.120M}, they used Doppler tomography and found $\mathrm{H_\alpha}$ emissions near the companion surface and the inner Lagrangian point. These $\mathrm{H_\alpha}$ emitting materials were suggested to be related to the intrabinary shock, and these regions could be a possible birthplace for the flares observed.}

\bb{
Based on the two-dimensional velocity Doppler map of the $\mathrm{H_\alpha}$ line, \cite{2021A&A...649A.120M} further claimed that the intrabinary shock is closer to the companion, contradicting to the X-ray observations (i.e., the shock is closer to the pulsar; \citealt{2018ApJ...866...71C,2019A&A...621L...9Y}). Consequently, they concluded that the X-ray and $\mathrm{H_\alpha}$ observations actually refer to the two sides of the shock that are closer to the pulsar and to the companion, respectively, indicating a very thick shock shell. Although a thin-shell approximation is often used in intrabinary shock models (e.g., \citealt{2020ApJ...904...91V}), a thick-shell shock is not entirely impossible. However, as discussed in \cite{2017ApJ...839...80W}, the accelerated, radiative electron population should retain in a relatively thin shell. We therefore speculate that the $\mathrm{H_\alpha}$ gases are likely unshocked intrabinary materials illuminated by the shock X-ray emission. This interpretation is also consistent with the low velocities of the emissions in the rotating frame of the binary (if the proposed positions are correct). In this case, the $\mathrm{H_\alpha}$ regions should not be powerful enough to radiate the observed flares with brightnesses comparable to the star light of the companion (i.e., the flare amplitude can be higher than 1~mag; Figure \ref{fig:j1048_lc}).
In contrast, the thin-shell shock with radiative electrons possibly generate the flares. However, this is unlikely in the case of \src\ as no X-ray eclipse was observed in the \textit{Chandra} X-ray observations (\citealt{2018ApJ...866...71C,2019A&A...621L...9Y}; Appendix \ref{app:chandra}), suggesting that the thin-shell shock was not completely blocked by the companion during the eclipses.}

Based on the irradiation power of $L_\mathrm{irr}=(1.9^{+0.4}_{-0.3})\times10^{33}$\lum\ inferred from the ELC modelling, the X-ray emission from the intrabinary shock ($L_\mathrm{x}=6.3\times10^{30}$\lum) is not powerful enough to heat up the companion as observed \citep{2017ApJ...845...42S,2018ApJ...866...71C,2019A&A...621L...9Y}. In contrast, the pulsar wind ($\dot{E}\approx10^{34}$\lum) as well as the $\gamma$-ray radiation ($L_\mathrm{\gamma}=2.9\times10^{32}$\lum) are possible energy sources for the irradiation \citep{1990ApJ...358..561H,2016ApJ...823..105D,2017ApJ...845...42S}. The observed flares can be a consequence of increased particle/photon fluxes from the pulsar that reach the companion surface \bb{through the intrabinary shock. In this context, the variable heating effect could result from unstable pulsar wind heating due to} the geometric change of the intrabinary shock possibly triggered by the magnetic activity of the companion \citep{2019A&A...621L...9Y}, although the detailed mechanism is still under debate. 

\subsection{Flare Eclipses v.s. Gamma-Ray Eclipses}

With \textit{Fermi}-LAT data, \cite{2023NatAs...7..451C} discovered $\gamma$-ray eclipses in 7 out of 49 redback and black widow millisecond pulsars, including \src. We note that the $\gamma$-ray eclipse duration of \src\ is $\theta_\mathrm{e,\gamma}=0.058$--$0.120$ orbits, which is significantly longer than the RIFT result. It could be the case that the evaporated wind causing the radio eclipses also attenuates the $\gamma$-ray emission \citep{2018MNRAS.476.1968P,2020ApJ...900..194K,2021ApJ...920...58K}. However, \cite{2023NatAs...7..451C} considered an isotropic model to prove the effect insignificant. In their calculation, the required column density is $\sim10^{25}$\cm\ \citep{1994ApJ...434..747P,2023NatAs...7..451C}, which is orders of magnitude higher than the canonical value of $\sim10^{18}$\cm\ inferred from the dispersion measures (DMs) around the radio elcipses \citep{1991ApJ...380..557R,2001MNRAS.321..576S,2017ApJ...839...80W,2020MNRAS.494.2948P}, suggesting the scenario unlikely.

Alternatively, the long $\gamma$-ray eclipse could indicate that some of the eclipsed $\gamma$-rays are from the apex region of the intrabinary shock located between the pulsar and the companion. It has been shown in several redback pulsars that the intrabinary shock emission can be comparable to the pulsar magnetosphere emission in the 0.1-1~GeV energy regime \citep{2014ApJ...797..111L,2018ApJ...863..194L,2018MNRAS.478.3987K}, in which \textit{Fermi}-LAT is most sensitive. Given that the pulsed and unpulsed $\gamma$-ray emission were both employed in the eclipse search for \src\ \citep{2023NatAs...7..451C}, the extended $\gamma$-ray eclipse observed could be the combined effect of the occultations of the pulsar and the intrabinary apex. In the above scenario, a simple geometrical calculation leads to 
\begin{equation}
R_*+r_s\approx(a-a_s)\sqrt{\cos^2 i\,\cos^2(\frac{\theta_\gamma}{2}) + \sin^2(\frac{\theta_\gamma}{2})},
\end{equation}
where $R_*$ is the radius of the companion, $r_s$ is the radius of the shock apex (a circular emission region is assumed), $a$ is the separation between the binary members, $a_s$ is distance to the apex from the pulsar, $i$ is the binary inclination, and $\theta_\gamma$ is the $\gamma$-ray eclipse duration in radians. We further assumed that $R_*\gg r_s$, given that the $\gamma$-ray emission falls/raise sharply during the eclipses \citep{2023NatAs...7..451C}. With $\theta_\gamma=0.36-0.75$ radians (equivalent to 0.058--0.12 orbits; \citealt{2023NatAs...7..451C}) and the best-fit ELC model parameters presented in Table \ref{tab:elc}, we find that $a_s\approx(1.4-6.1)\times10^{10}\,$cm. With the apex location, we inferred the momentum flux ratio between the winds from the companion and the pulsar by (e.g., \citealt{2014ApJ...797..111L})
\begin{equation}
\eta\approx(\frac{a}{r_s}-1)^\frac{1}{2}.
\end{equation}
It gives $\eta\approx2-98$, indicating that the stellar wind wins the pulsar wind and pushes the shock to wrap the pulsar. 
Interestingly, the shock opening direction affects the X-ray orbital modulation in redback systems. Due to the Doppler boosting effect, the X-ray emission is enhanced when the companion is behind the pulsar \citep{2014ApJ...797..111L} in the case of $\eta>1$, and this is consistent with the \textit{Chandra} observations of \src\ \citep{2018ApJ...866...71C,2019A&A...621L...9Y}. A larger-than-unity momentum flux ratio seems counter-intuitive, given that the pulsar wind is relativistic. However, by considering the rapid rotation of the companion with the dynamo effect, the stellar wind can be powerful with the help of the strong magnetic pressure. In PSR~J1023+0038, $\eta>1$ was also detected, and this was attributed to the strong magnetic pressure \citep{2013arXiv1311.5161A}. In this context, the momentum flux ratio \citep{2014ApJ...797..111L} can be written as
\begin{equation}
\eta_b\approx \frac{B_*^2\,R_*^2\,c}{L_\mathrm{sd}},
\end{equation}
where $B_*$ is the magnetic field strength at the surface of the companion, $c$ is the speed of light, and $L_\mathrm{sd}$ is the spin-down power of the pulsar, which is $L_\mathrm{sd}=1.2\times10^{34}$\lum\ for \src\ \citep{2016ApJ...823..105D,2021ApJ...909....6D}. \bb{Using the inferred $\eta$ parameter range, the magnetic field strength is approximately $B_*\approx30-170\,$G, which is higher than the typical strengths of mature K-type main-sequences (e.g., $2-17\,$G for 61~Cygni~A and $\sim2.5\,$G for HD~219134; \citealt{2016A&A...594A..29B,2018MNRAS.481.5286F}).}

\subsection{A New Tool for Pulsar Mass Measurement}

The discovery of the flare eclipses of \src\ and its application to mass measurement opens a new pathway for astronomers to enrich the demographics of neutron stars. Currently, the most accurate mass measurements for neutron stars rely on the general relativistic effects of massive systems, such as Shapiro delay and the rates of the periastron advances \citep{2003Natur.426..531B,2010Natur.467.1081D,2010ApJ...722.1030W,2014MNRAS.443.2183F,2015ApJ...812..143M,2016ApJ...832..167F,2018ApJ...854L..22S,2020NatAs...4...72C}. Besides, high-resolution spectroscopy of the companions can make good constraints on the companion masses or orbital inclinations for fine pulsar mass measurements \citep{2013Sci...340..448A,2022MNRAS.512.3001K}. Despite the remarkable results, competitive observing times from big telescopes are essential. By contrast, our investigation on \src\ has demonstrated a much more accessible approach using a small optical telescope. Flaring episodes are not uncommon in redbacks and black widows \citep{2017ApJ...850..100A,2018ApJ...866...71C,2020ApJ...895L..36P,2022ApJ...935..151H,2022ApJ...941..199S}, although some have not exhibited yet, perhaps owing to a lack of continues optical observations. With high-cadence long-term monitoring even with small aperture telescopes, it is convinced that the flare eclipse measurement can be transferable to other spider systems that are viewed sufficiently edge-on. The immediate candidates would be the seven systems that have shown $\gamma$-ray eclipses in the \textit{Fermi} Large Area Telescope (LAT) data \citep{2023NatAs...7..451C}. The possible differences between the $\gamma$-ray and optical eclipse durations can also be a probe for the location of the shock apex as well as the magnetic field strength of the spider companion. Moreover, the data of some large optical survey programs, like the Rubin Observatory Legacy Survey of Space and Time (LSST), would be useful on the mass measurements for faint spiders. With more than eighty spider known systems today \citep{2019Galax...7...93H,2025arXiv250511691K}, improved mass measures of some of them can already directly enhance our understanding of neutron stars.

\begin{acknowledgements}
This work used high-performance computing facilities operated by the Center for Informatics and Computation in Astronomy (CICA) at National Tsing Hua University. This equipment was funded by the Ministry of Education of Taiwan, the National Science and Technology Council of Taiwan, and National Tsing Hua University.
This research has made use of data obtained from the \textit{Chandra} Data Archive and the \textit{Chandra} Source Catalog, and software provided by the \textit{Chandra} X-ray Center (CXC) in the application packages CIAO and Sherpa.

K.L.L. and K.Y.A. are supported by the National Science and Technology Council of the Republic of China (Taiwan) through grants NSTC 113-2636-M-006-003 and NSTC 114-2112-M-006-035.
This research is partially supported by the Yushan Fellow Program by the Ministry of Education (MOE), Taiwan (MOE-114-YSFMS-0005-002-P2).
L.C.C.L. is supported by the National Science and Technology Council of the Republic of China (Taiwan) through grant NSTC 114-2811-M-006-047-MY3.
A.K.H.K. is supported by the National Science and Technology Council of the Republic of China (Taiwan) through grants 113-2112-M-007-001 and 114-2112-M-007-033-MY3.
C.Y.H. is supported by the National Research Foundation of Korea grant RS-2025-16070477.
J.T. is supported by the National Key Research and Development Program of China (grant No. 2020YFC2201400) and the National Natural Science Foundation of China (grant No. 12173014).
\end{acknowledgements}
\facility{CXO}

\clearpage
\appendix
\restartappendixnumbering
\setcounter{figure}{6}

\section{Individual RIFT light curves}
\label{app:rift}

\begin{figure}[ht]
	\begin{subfigure}{}
	\includegraphics[width=0.44\textwidth]{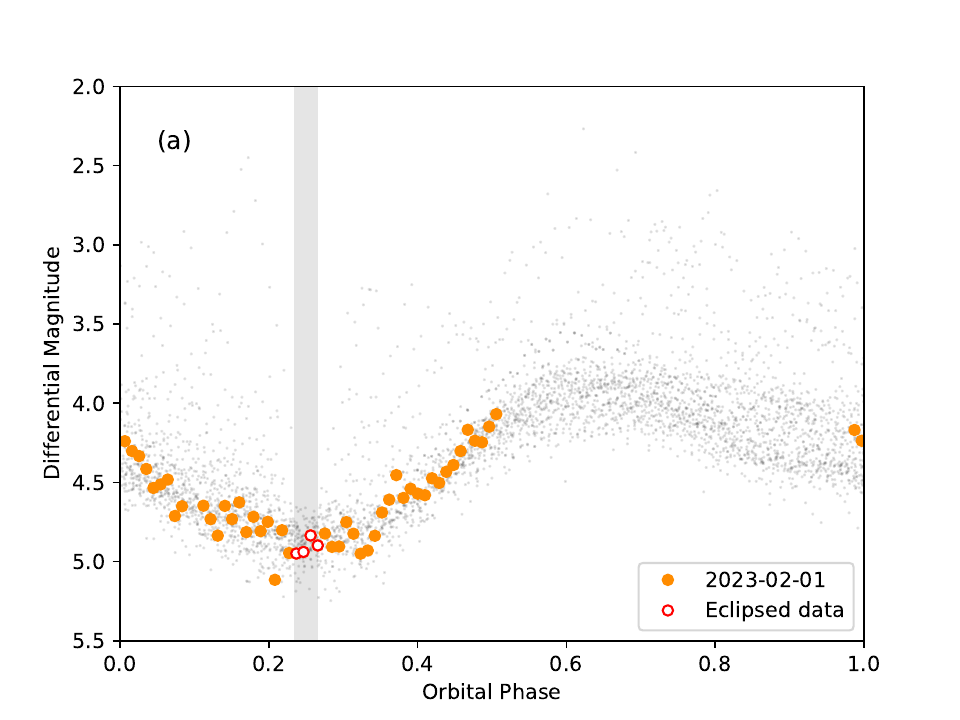}
	\end{subfigure}
	\begin{subfigure}{}
	\includegraphics[width=0.44\textwidth]{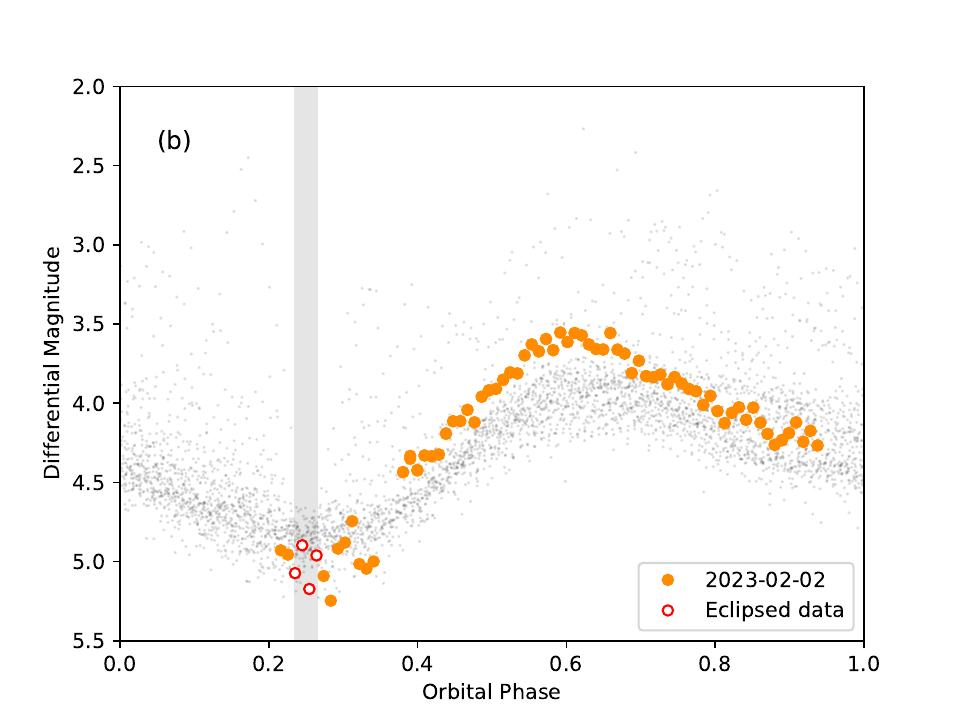}
	\end{subfigure}
	\begin{subfigure}{}
	\includegraphics[width=0.44\textwidth]{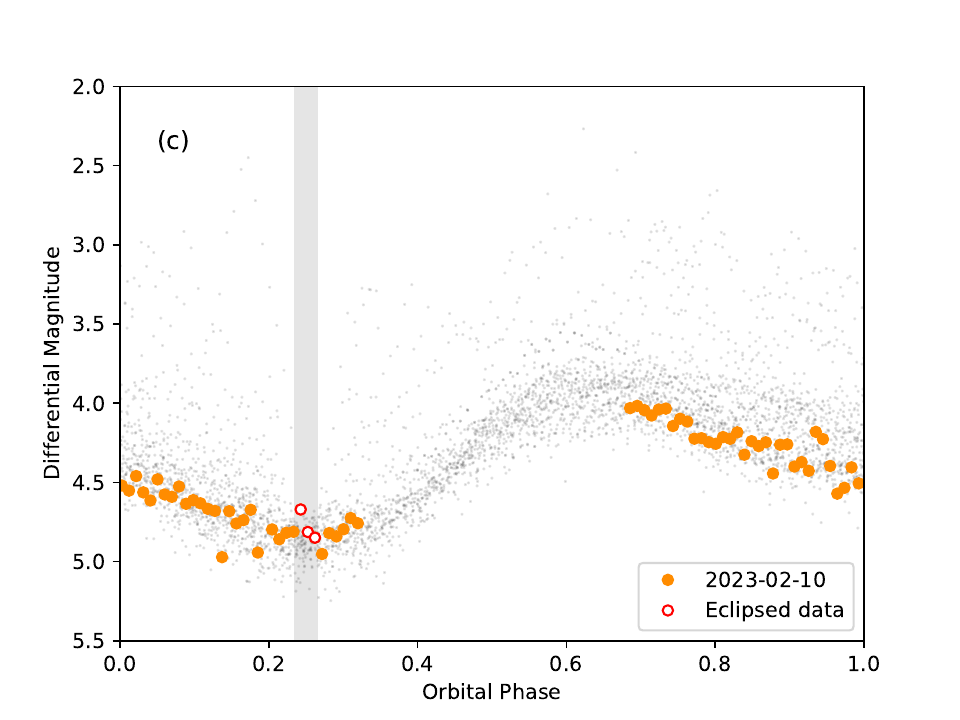}
	\end{subfigure}
	\begin{subfigure}{}
	\includegraphics[width=0.44\textwidth]{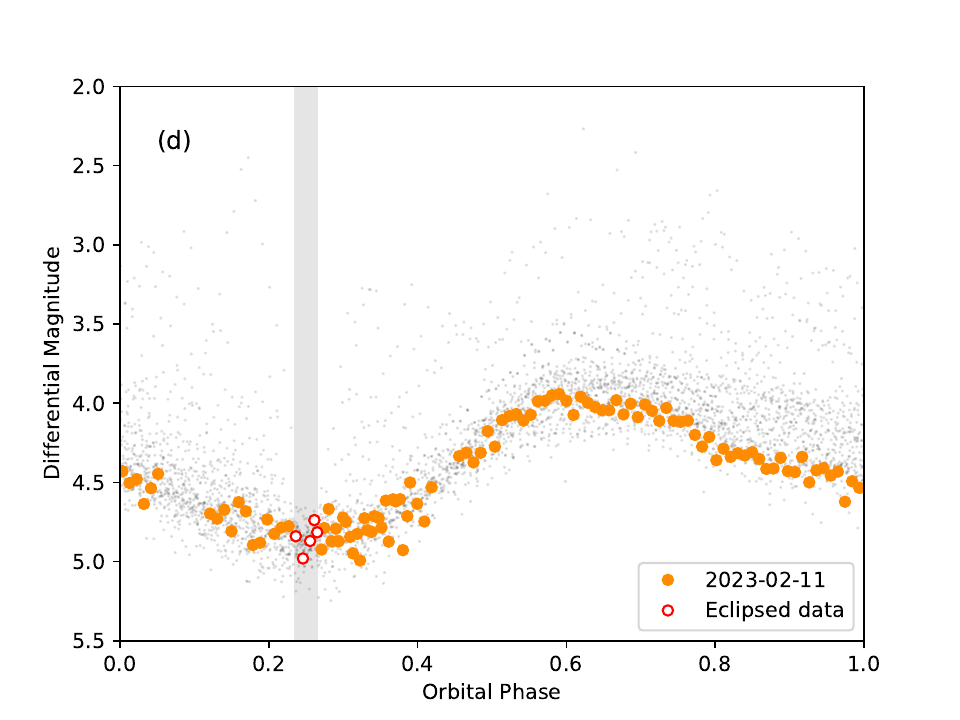}
	\end{subfigure}
	\begin{subfigure}{}
	\includegraphics[width=0.44\textwidth]{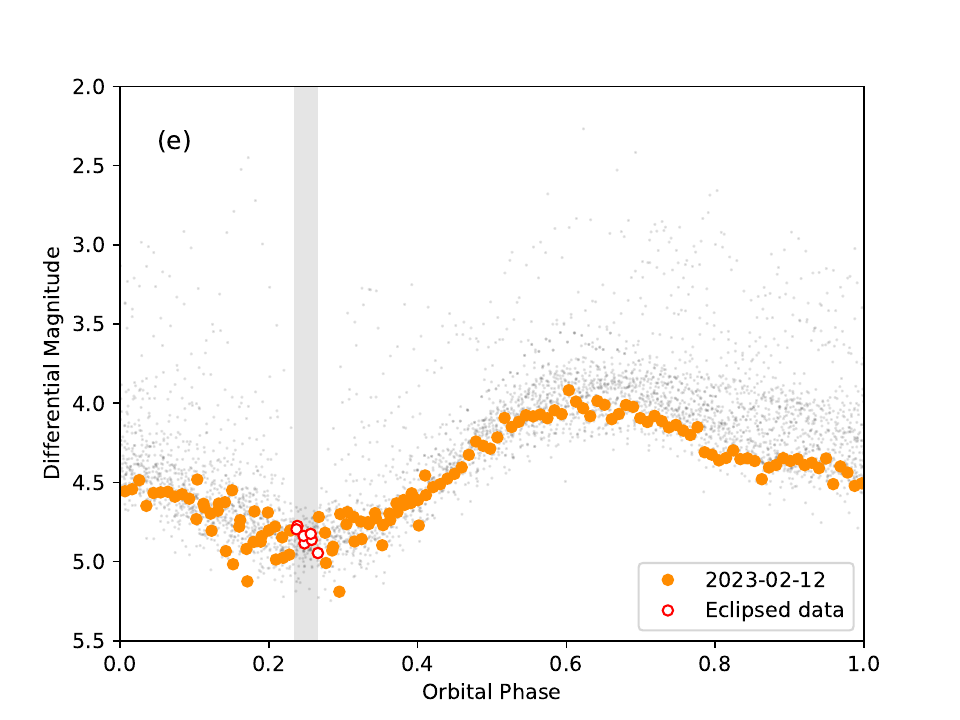}
	\end{subfigure}
	\begin{subfigure}{}
	\includegraphics[width=0.44\textwidth]{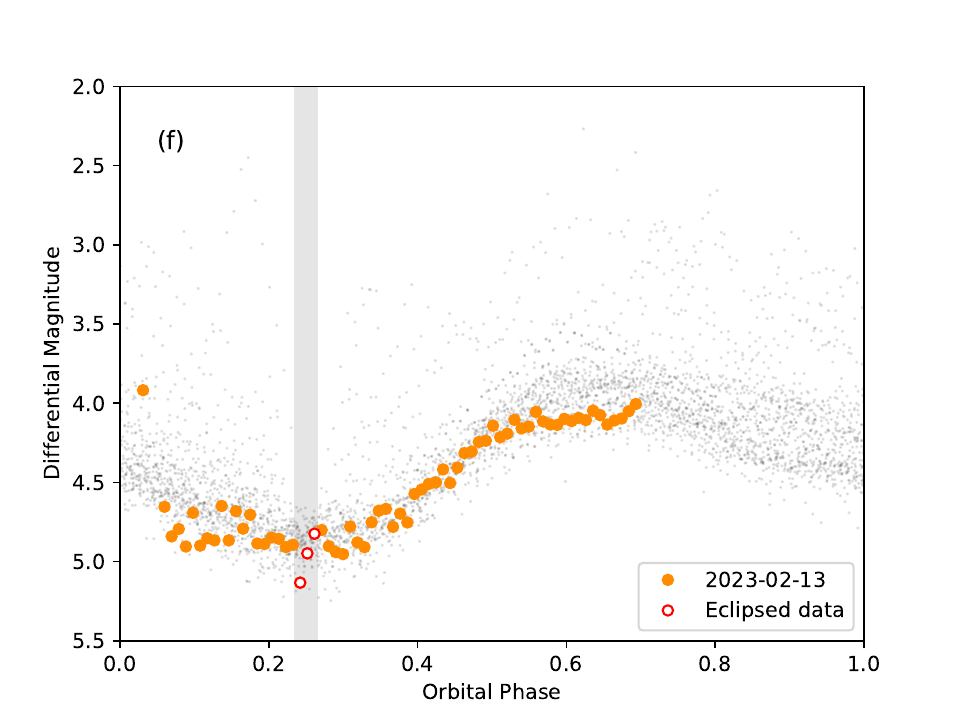}
	\end{subfigure}
\centering
\caption{The phased light curves of \src\ of each night. The plots are similar to Figure \ref{fig:j1048_lc}, but the k-means clustering classifications are not indicated. Instead, the flares identified visually are marked dark dots at the centers of the data and the eclipsed data are shown as red hollow circles. The grey dots are the stacked light curve of all the RIFT observations for comparison.}
\label{fig:individual_obs}
\end{figure}

\setcounter{figure}{6}

\begin{figure}
	\begin{subfigure}{}
	\includegraphics[width=0.44\textwidth]{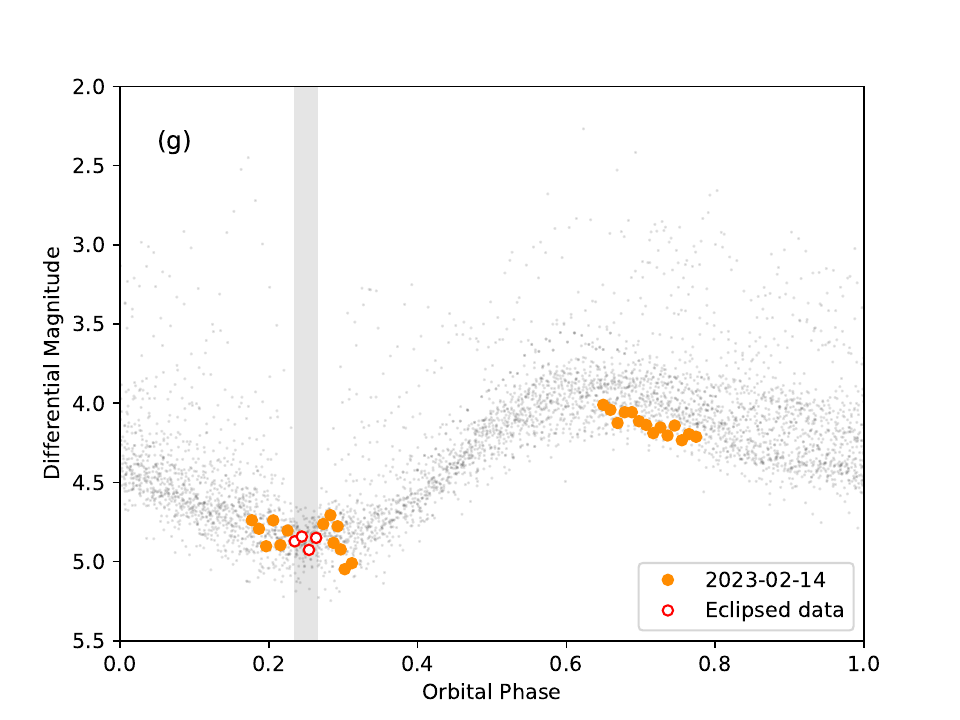}
	\end{subfigure}
	\begin{subfigure}{}
	\includegraphics[width=0.44\textwidth]{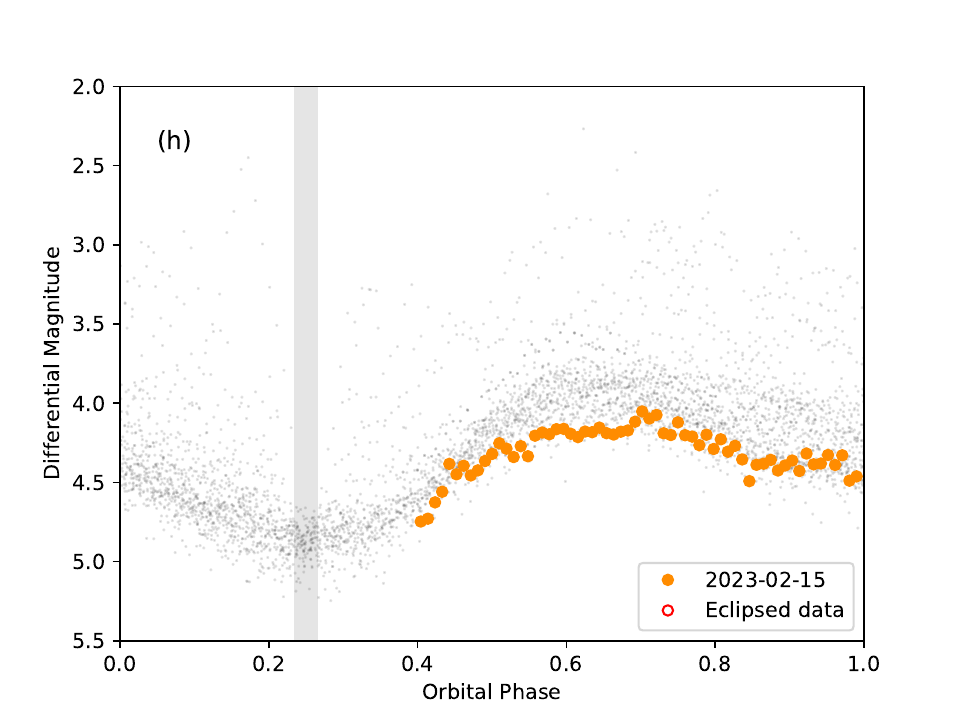}
	\end{subfigure}
	\begin{subfigure}{}
	\includegraphics[width=0.44\textwidth]{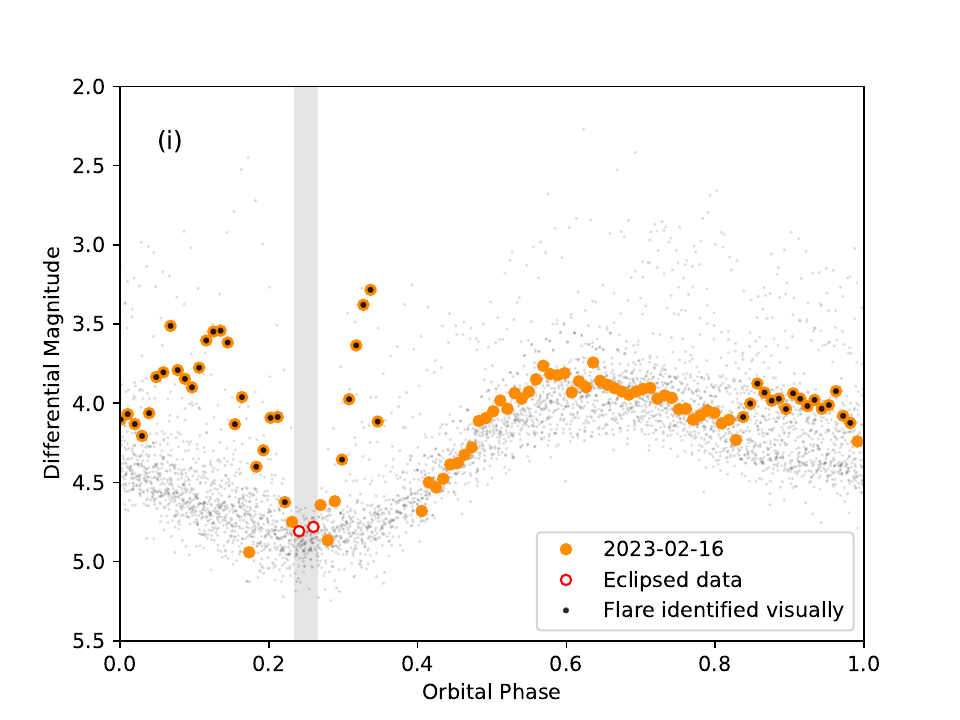}
	\end{subfigure}
	\begin{subfigure}{}
	\includegraphics[width=0.44\textwidth]{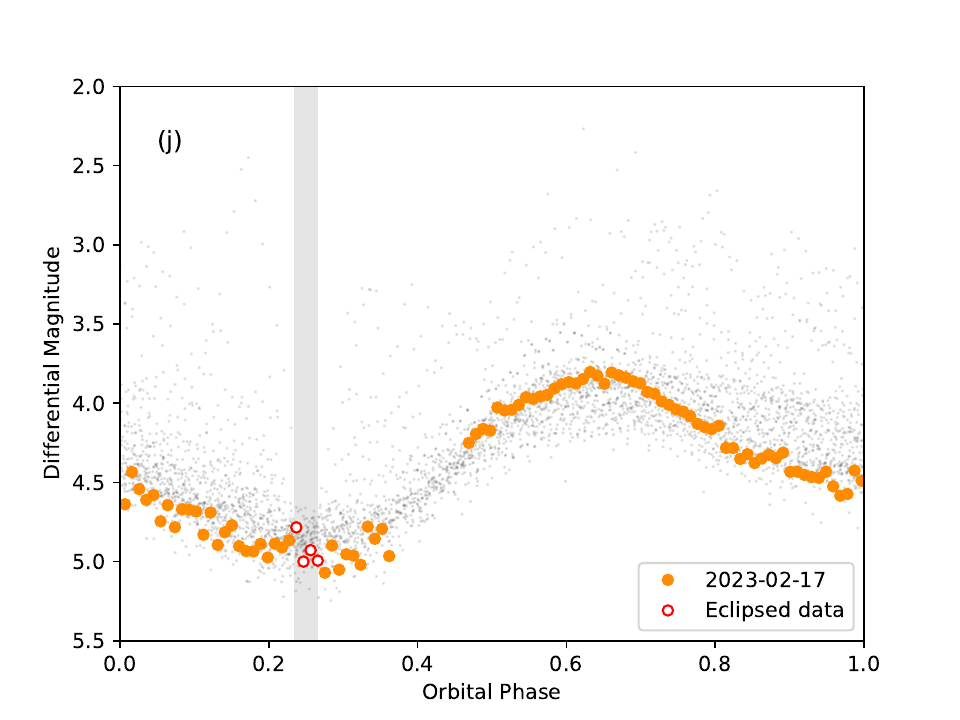}
	\end{subfigure}
	\begin{subfigure}{}
	\includegraphics[width=0.44\textwidth]{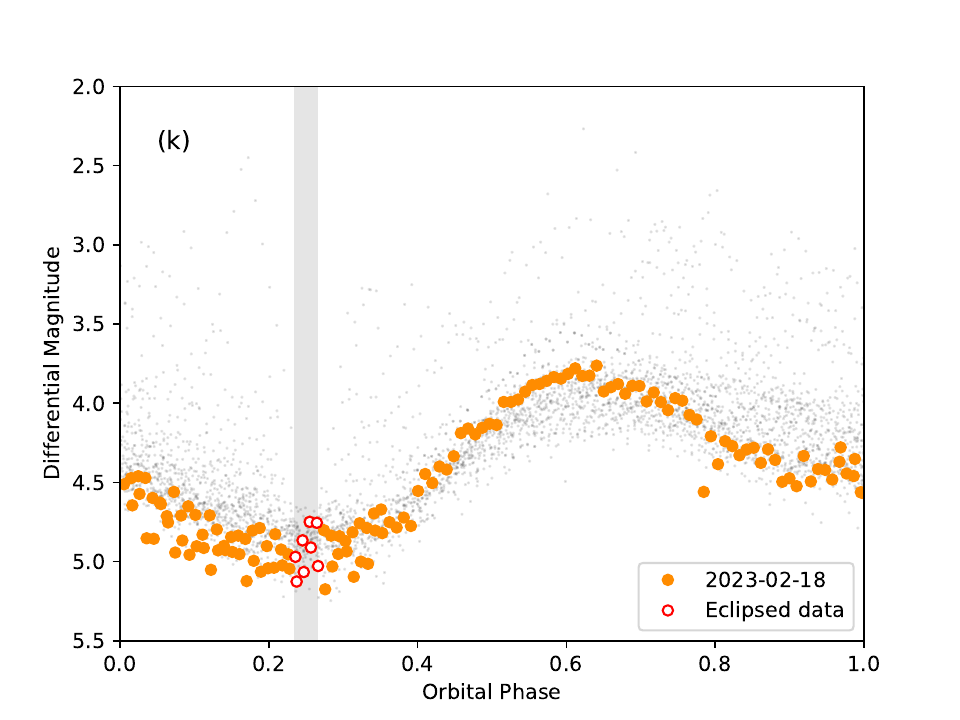}
	\end{subfigure}
	\begin{subfigure}{}
	\includegraphics[width=0.44\textwidth]{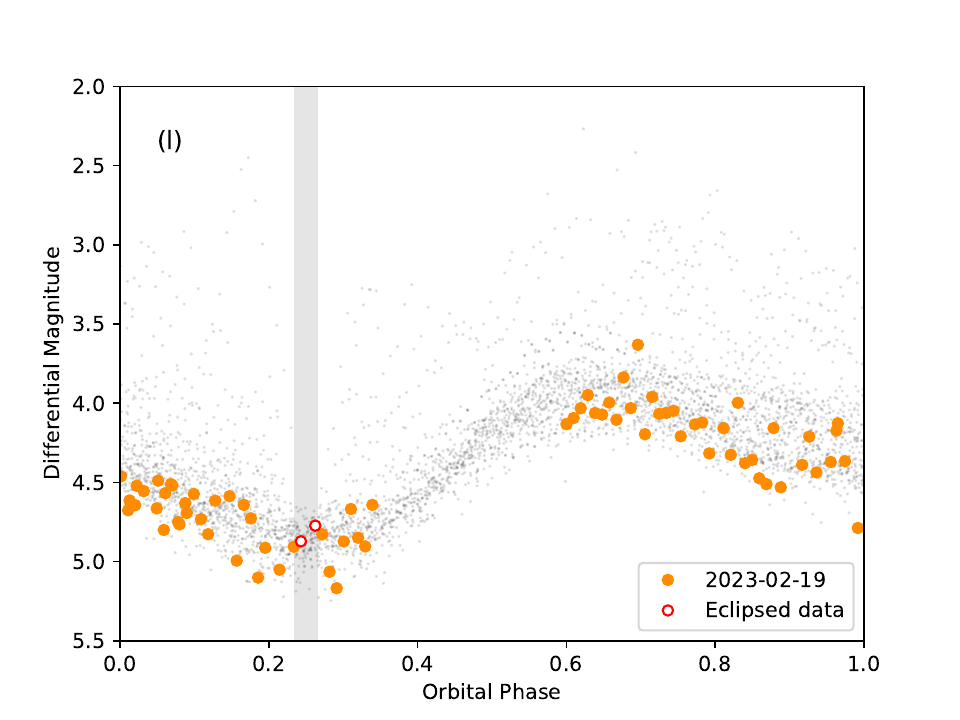}
	\end{subfigure}
\centering
\caption{\textit{continued}}
\end{figure}

\setcounter{figure}{6}

\begin{figure}
	\begin{subfigure}{}
	\includegraphics[width=0.44\textwidth]{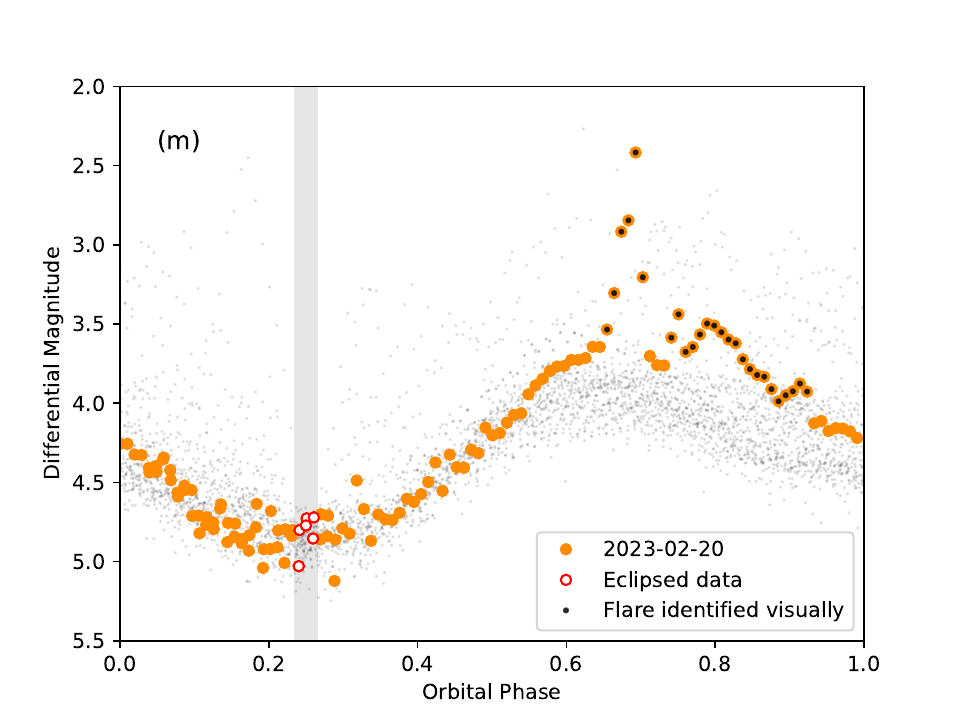}
	\end{subfigure}
	\begin{subfigure}{}
	\includegraphics[width=0.44\textwidth]{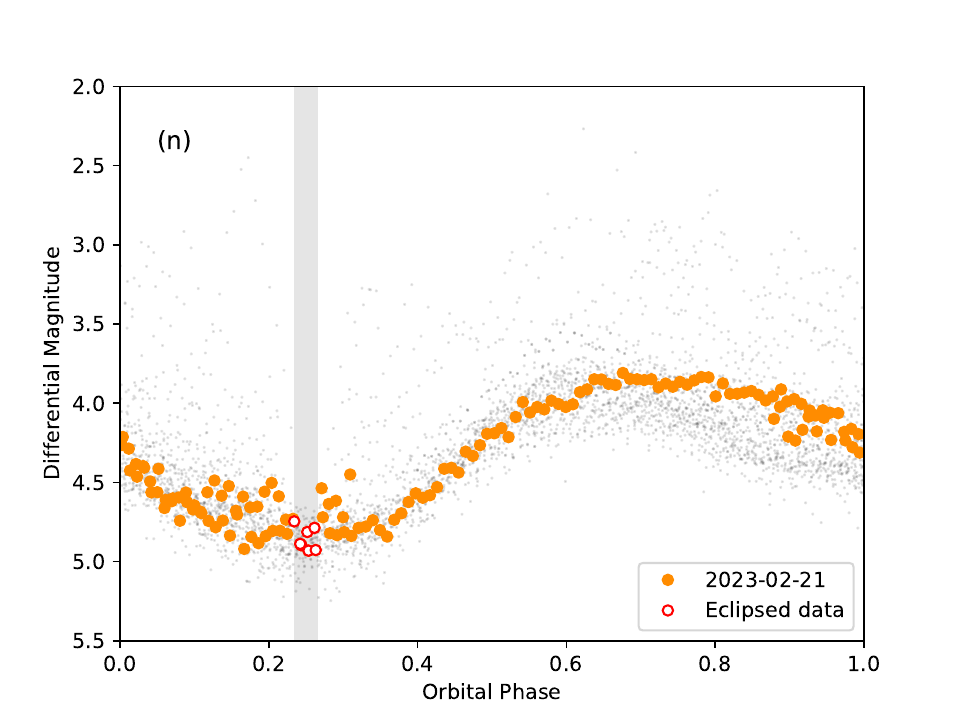}
	\end{subfigure}
	\begin{subfigure}{}
	\includegraphics[width=0.44\textwidth]{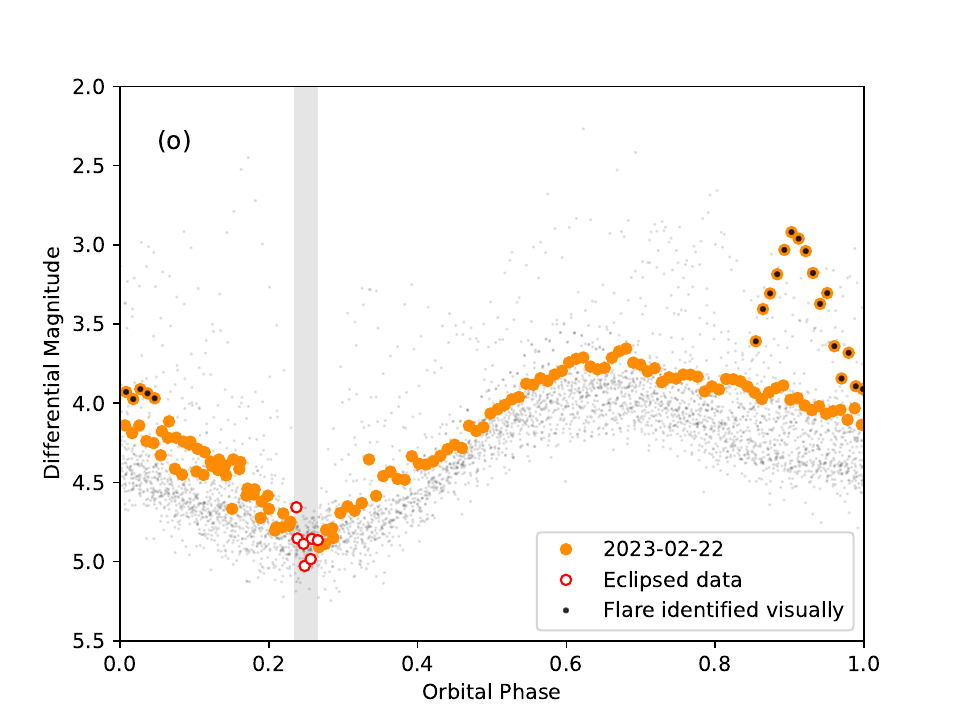}
	\end{subfigure}
	\begin{subfigure}{}
	\includegraphics[width=0.44\textwidth]{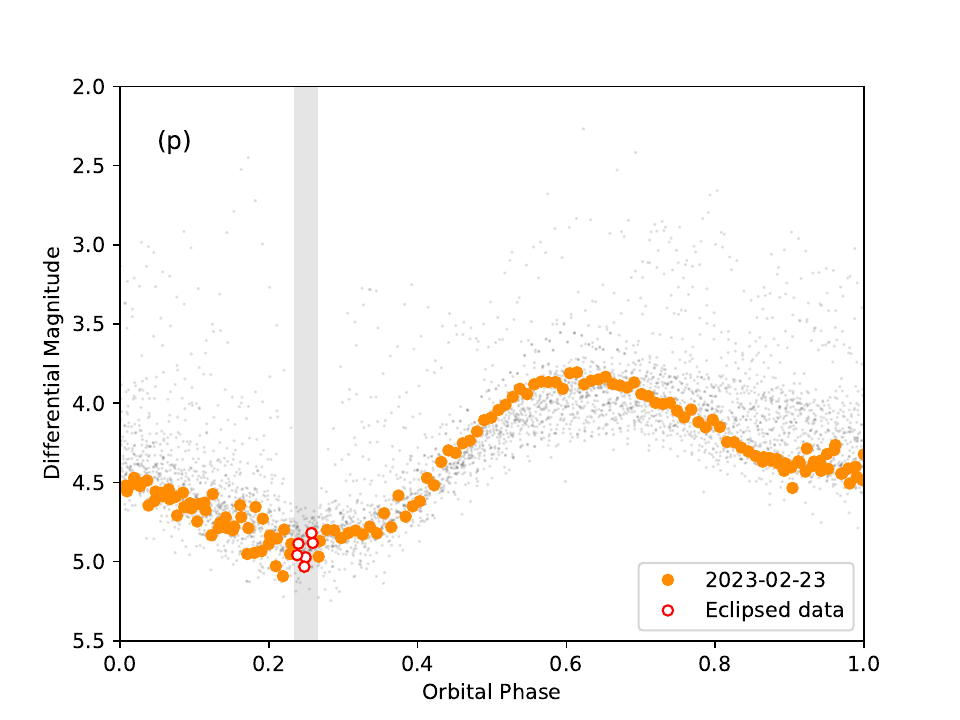}
	\end{subfigure}
	\begin{subfigure}{}
	\includegraphics[width=0.44\textwidth]{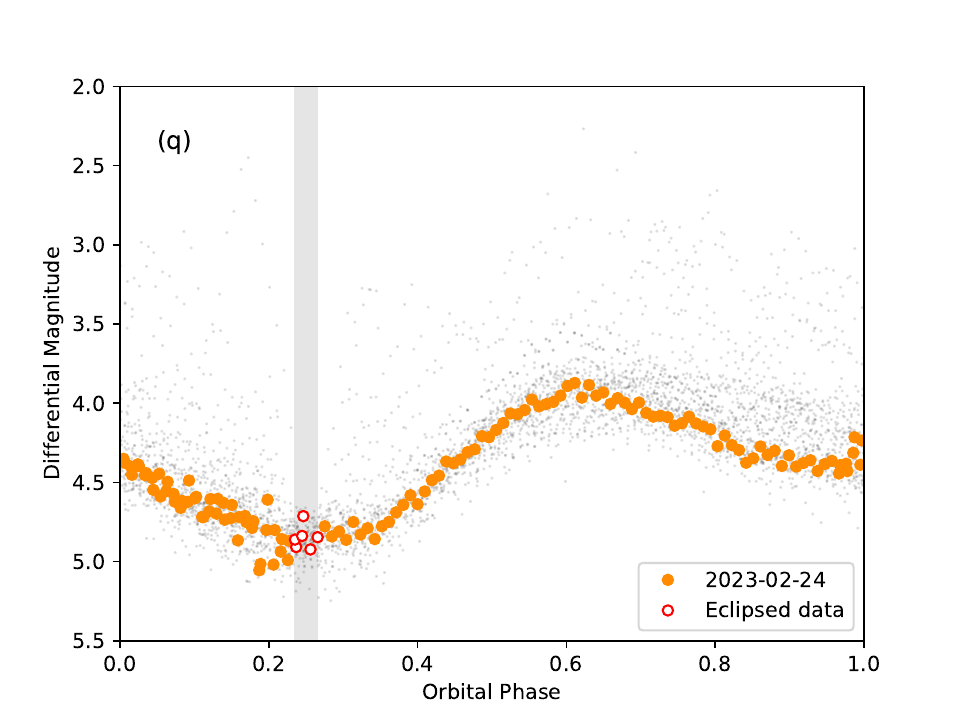}
	\end{subfigure}
	\begin{subfigure}{}
	\includegraphics[width=0.44\textwidth]{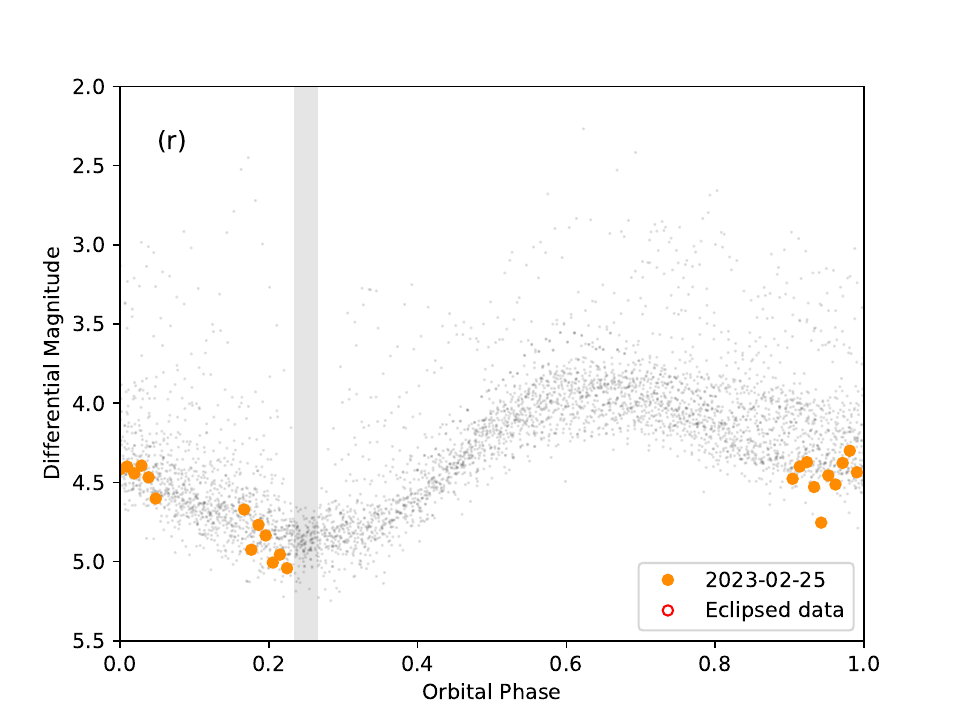}
	\end{subfigure}
\centering
\caption{\textit{continued}}
\end{figure}

\setcounter{figure}{6}

\begin{figure}
	\begin{subfigure}{}
	\includegraphics[width=0.44\textwidth]{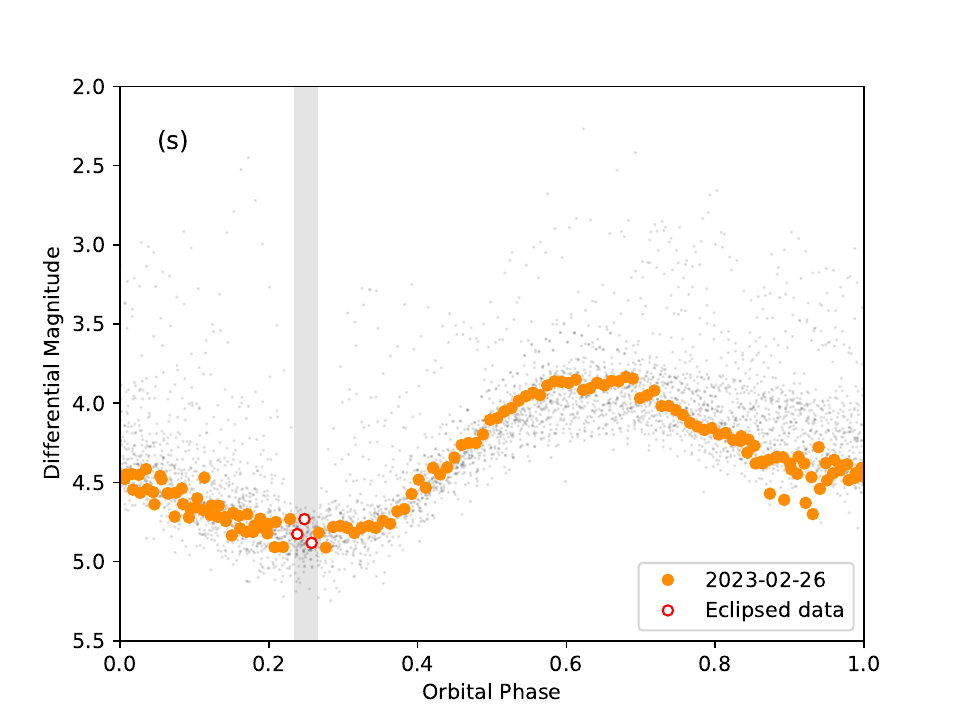}
	\end{subfigure}
	\begin{subfigure}{}
	\includegraphics[width=0.44\textwidth]{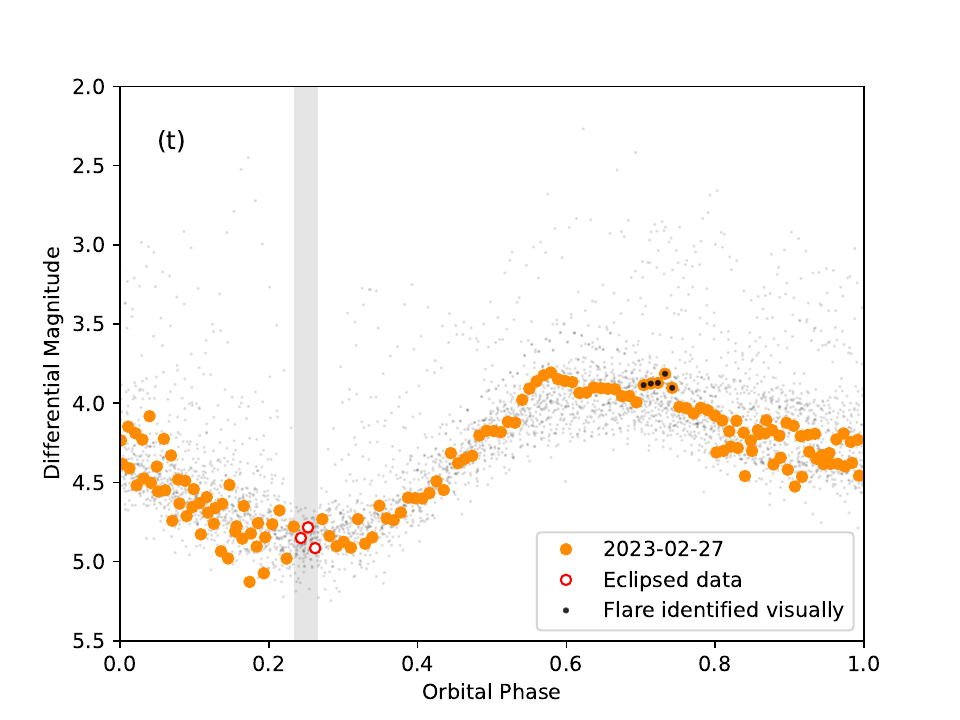}
	\end{subfigure}
	\begin{subfigure}{}
	\includegraphics[width=0.44\textwidth]{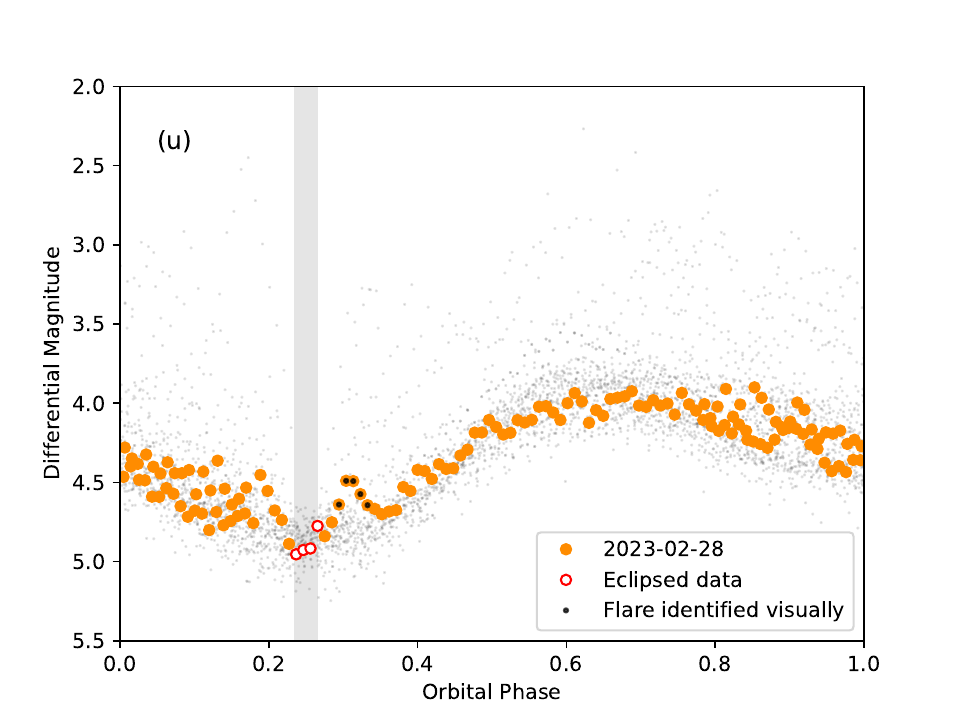}
	\end{subfigure}
	\begin{subfigure}{}
	\includegraphics[width=0.44\textwidth]{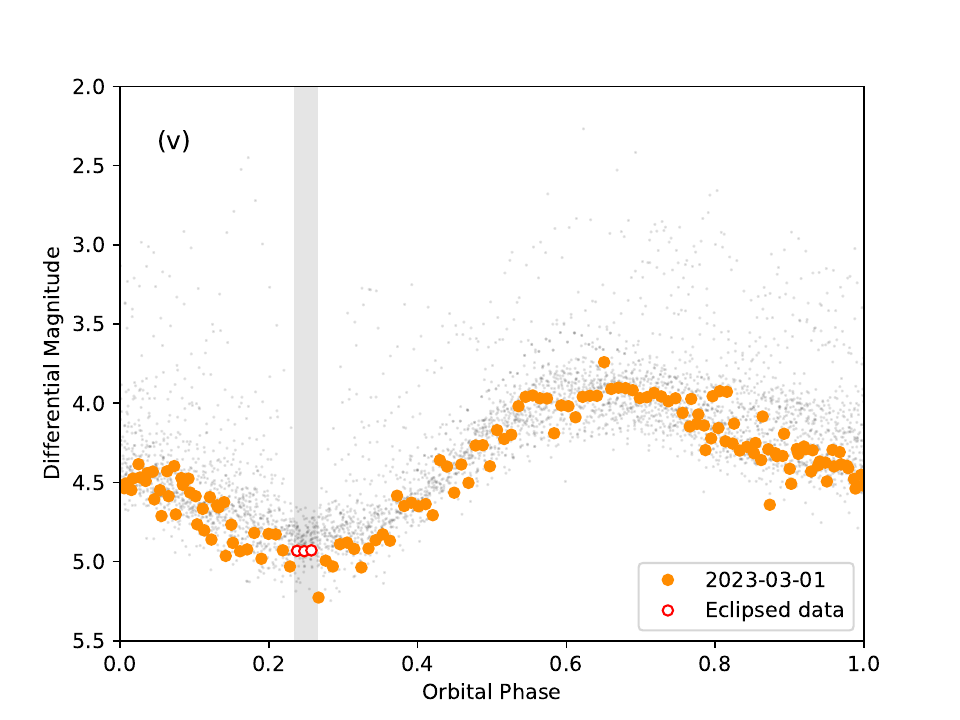}
	\end{subfigure}
	\begin{subfigure}{}
	\includegraphics[width=0.44\textwidth]{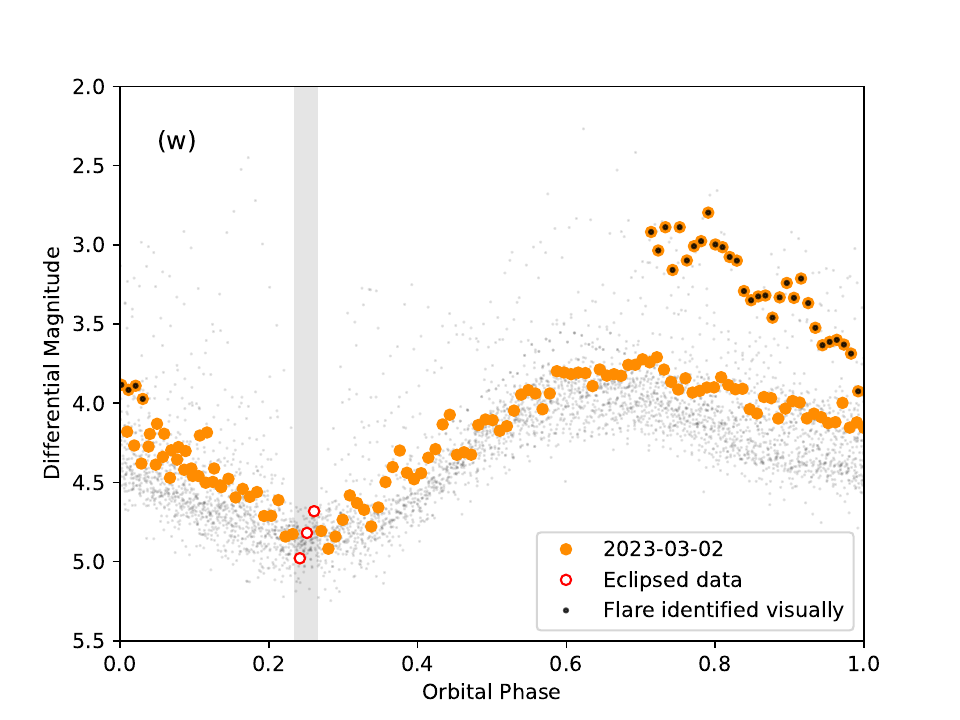}
	\end{subfigure}
	\begin{subfigure}{}
	\includegraphics[width=0.44\textwidth]{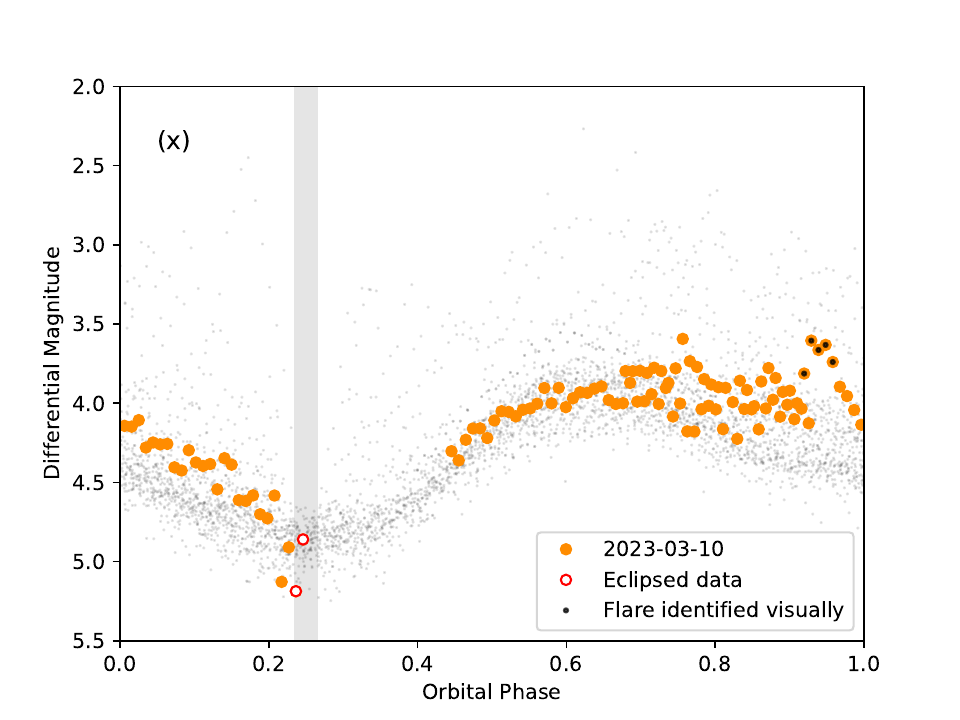}
	\end{subfigure}
\centering
\caption{\textit{continued}}
\end{figure}

\setcounter{figure}{6}

\begin{figure}
	\begin{subfigure}{}
	\includegraphics[width=0.44\textwidth]{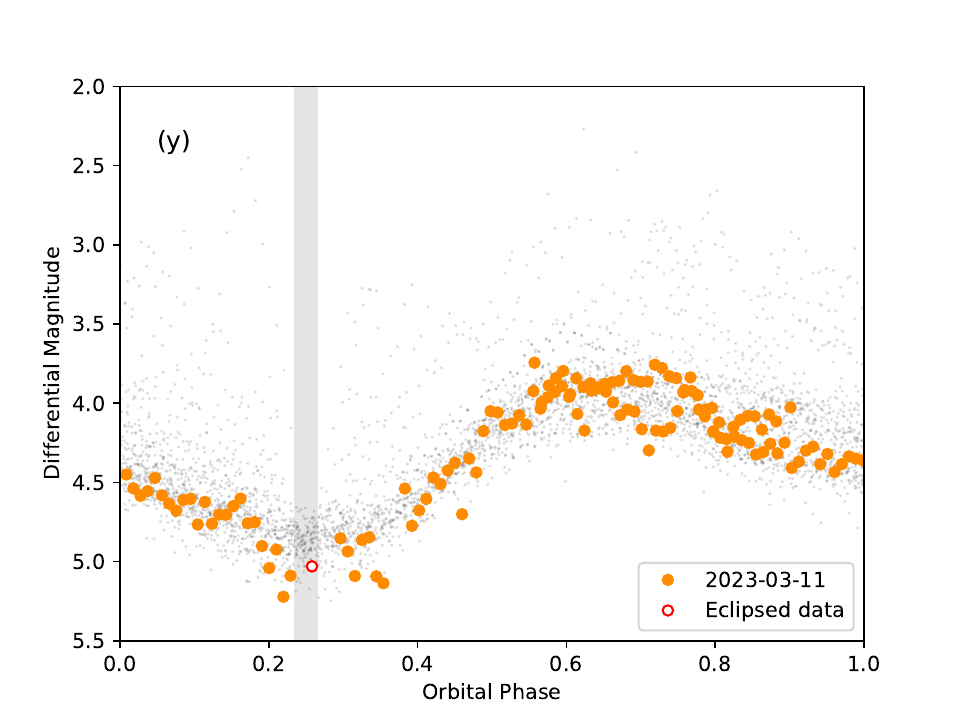}
	\end{subfigure}
	\begin{subfigure}{}
	\includegraphics[width=0.44\textwidth]{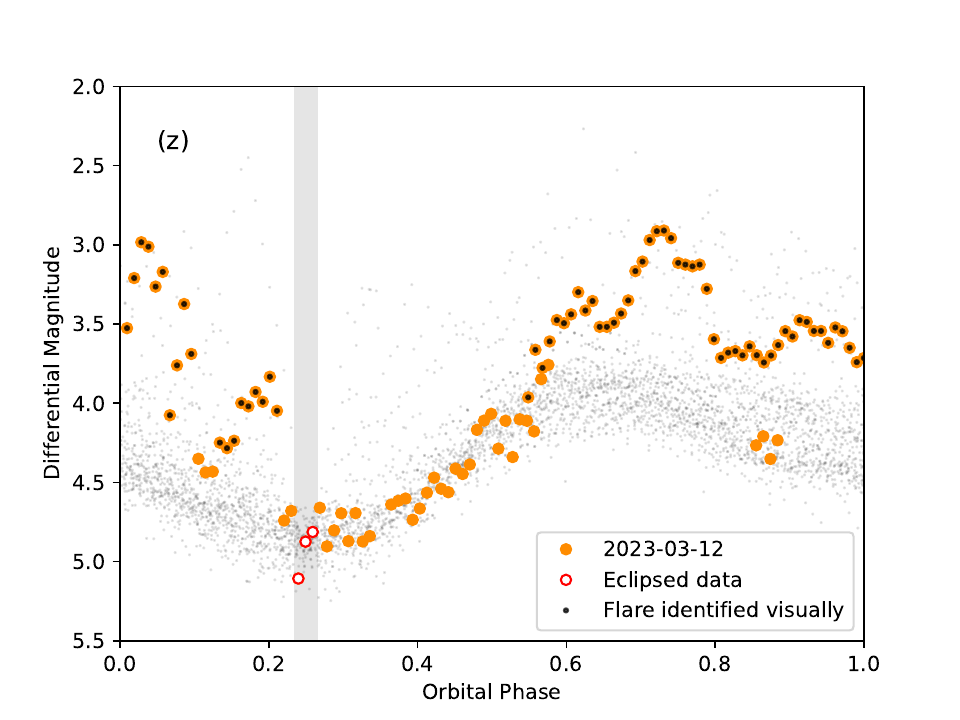}
	\end{subfigure}
	\begin{subfigure}{}
	\includegraphics[width=0.44\textwidth]{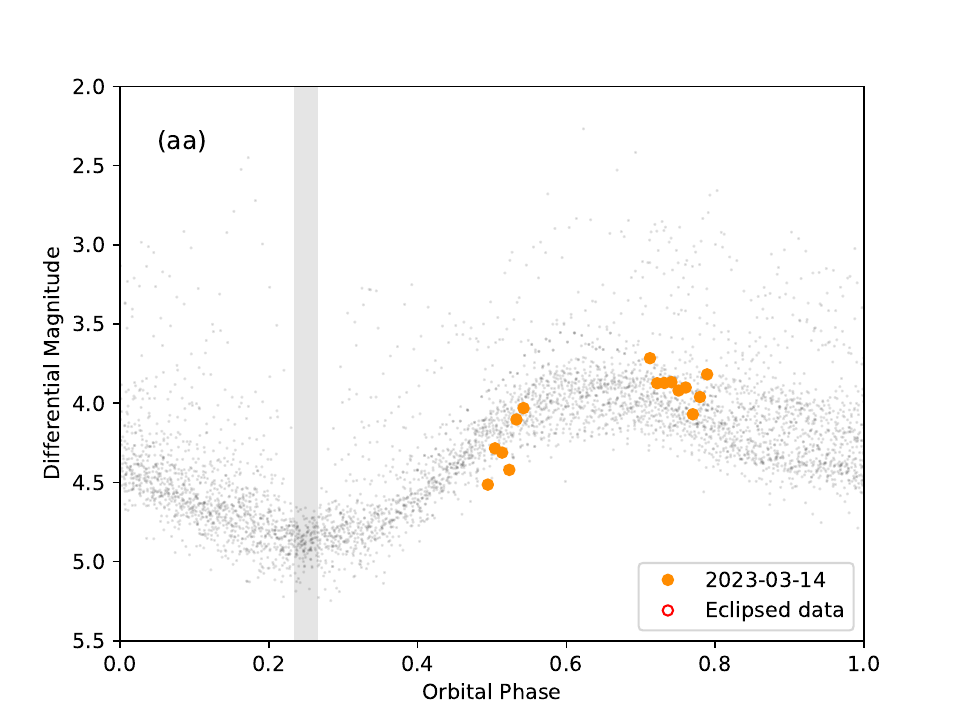}
	\end{subfigure}
	\begin{subfigure}{}
	\includegraphics[width=0.44\textwidth]{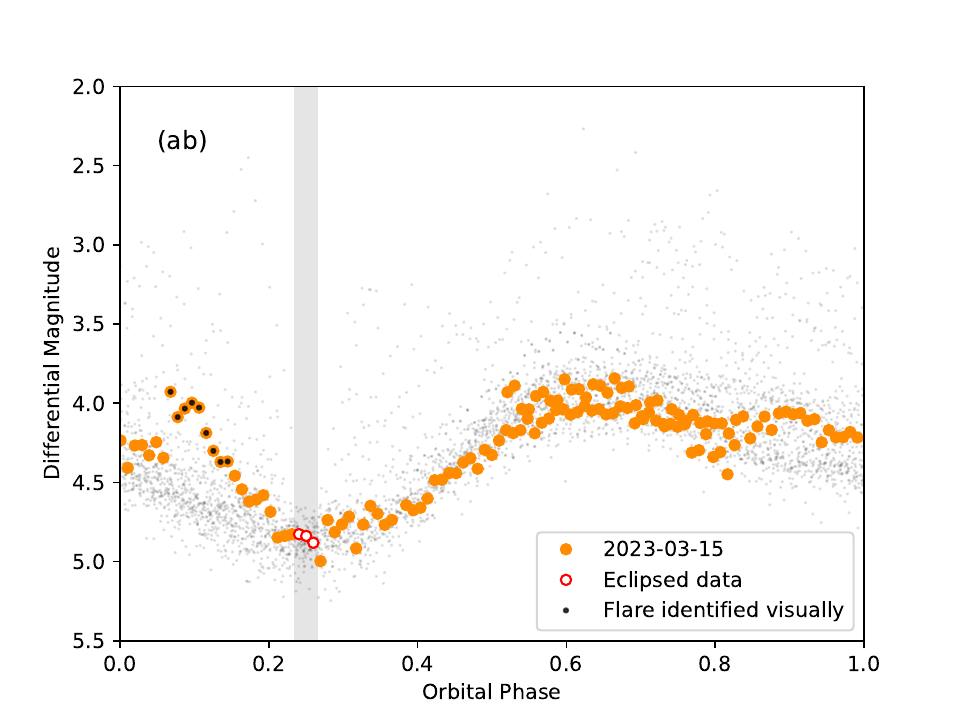}
	\end{subfigure}
	\begin{subfigure}{}
	\includegraphics[width=0.44\textwidth]{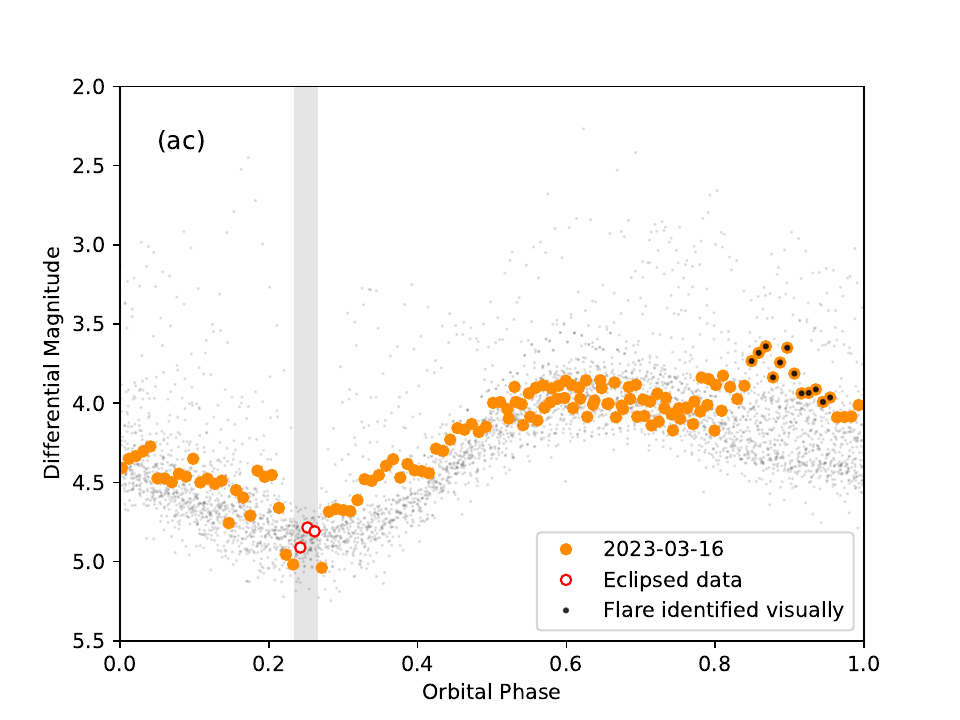}
	\end{subfigure}
	\begin{subfigure}{}
	\includegraphics[width=0.44\textwidth]{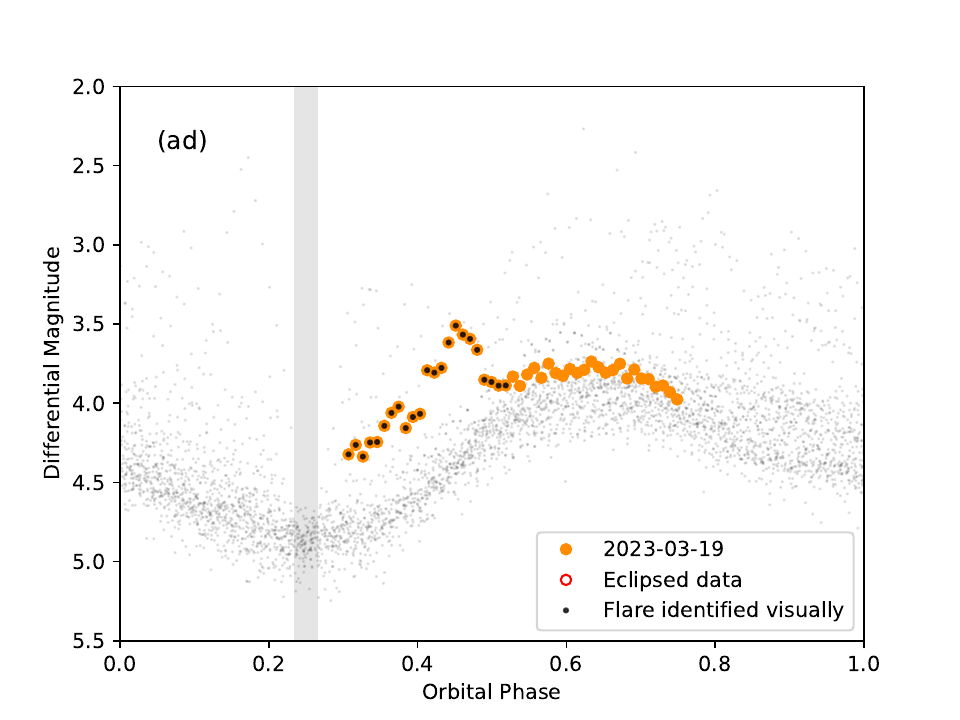}
	\end{subfigure}
\centering
\caption{\textit{continued}}
\end{figure}

\setcounter{figure}{6}

\begin{figure}
	\begin{subfigure}{}
	\includegraphics[width=0.44\textwidth]{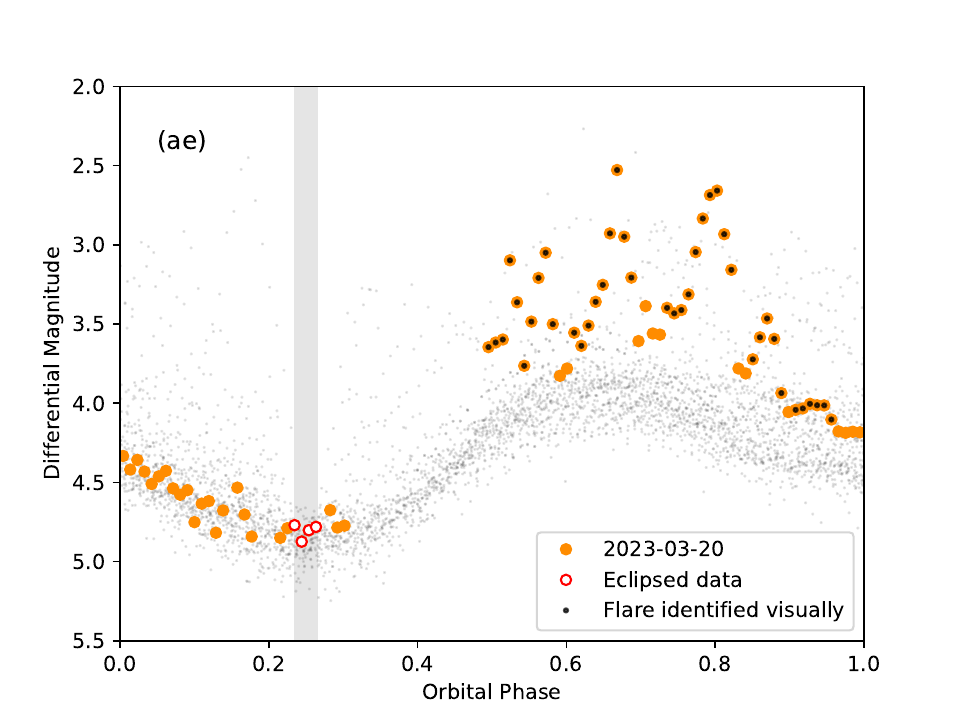}
	\end{subfigure}
	\begin{subfigure}{}
	\includegraphics[width=0.44\textwidth]{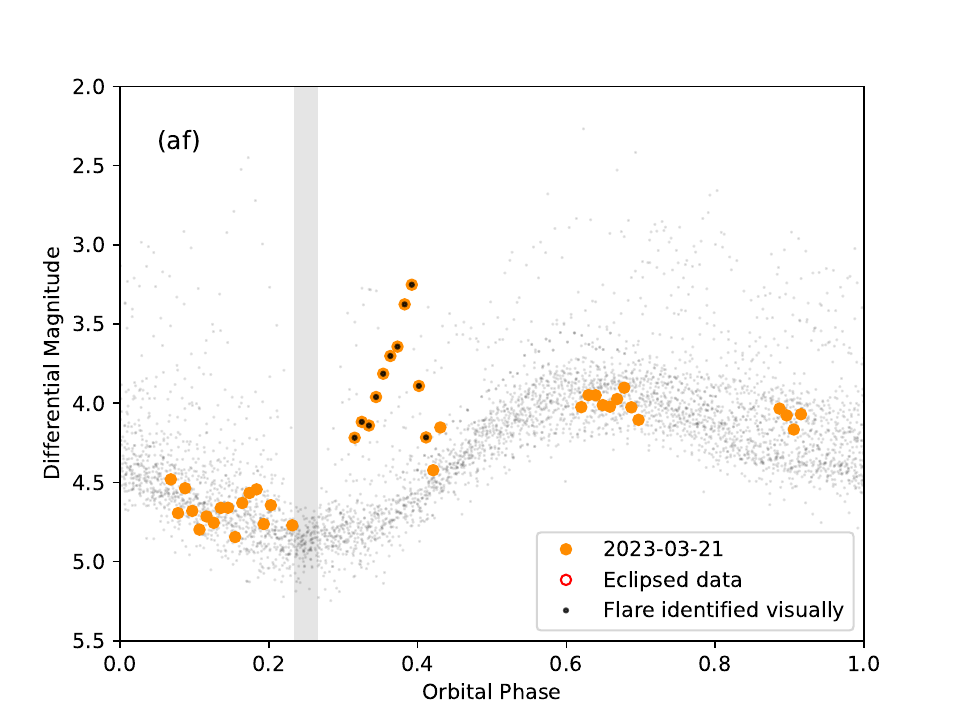}
	\end{subfigure}
	\begin{subfigure}{}
	\includegraphics[width=0.44\textwidth]{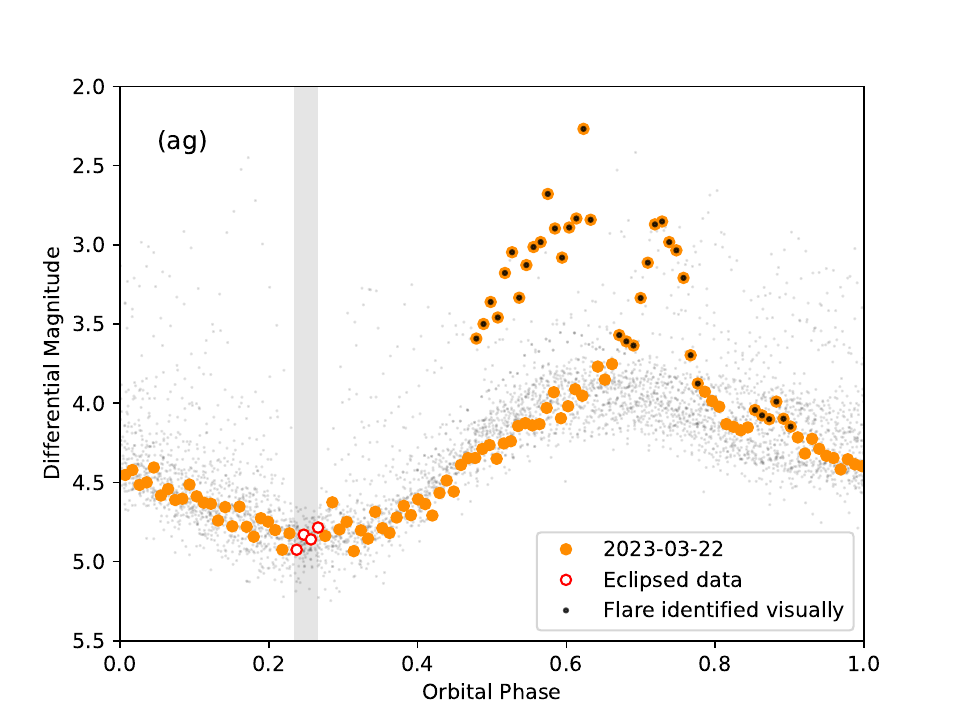}
	\end{subfigure}
	\begin{subfigure}{}
	\includegraphics[width=0.44\textwidth]{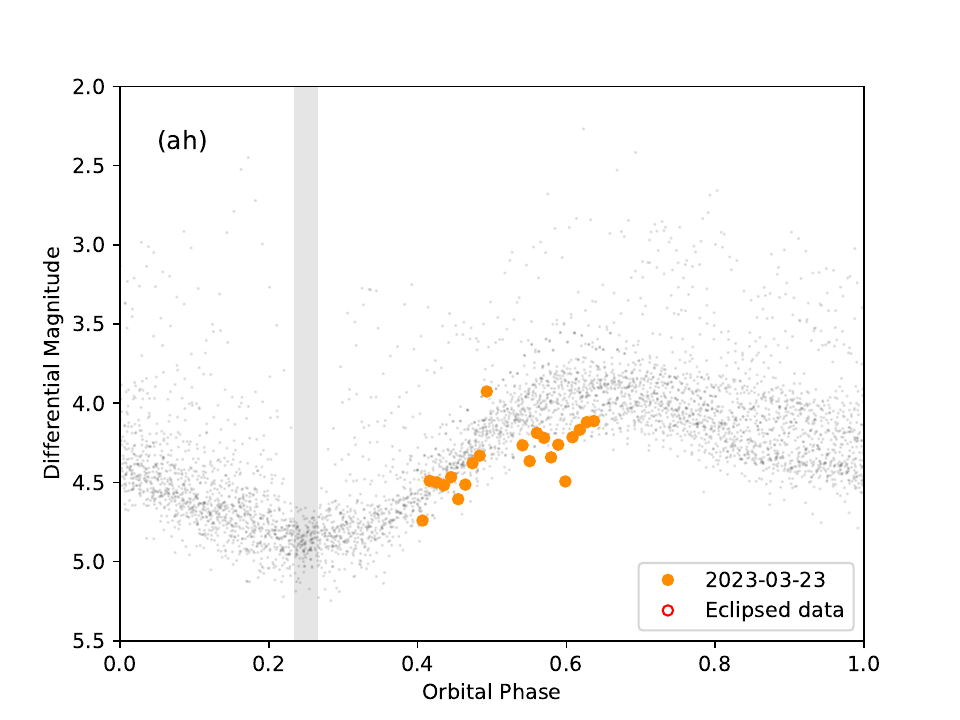}
	\end{subfigure}
	\begin{subfigure}{}
	\includegraphics[width=0.44\textwidth]{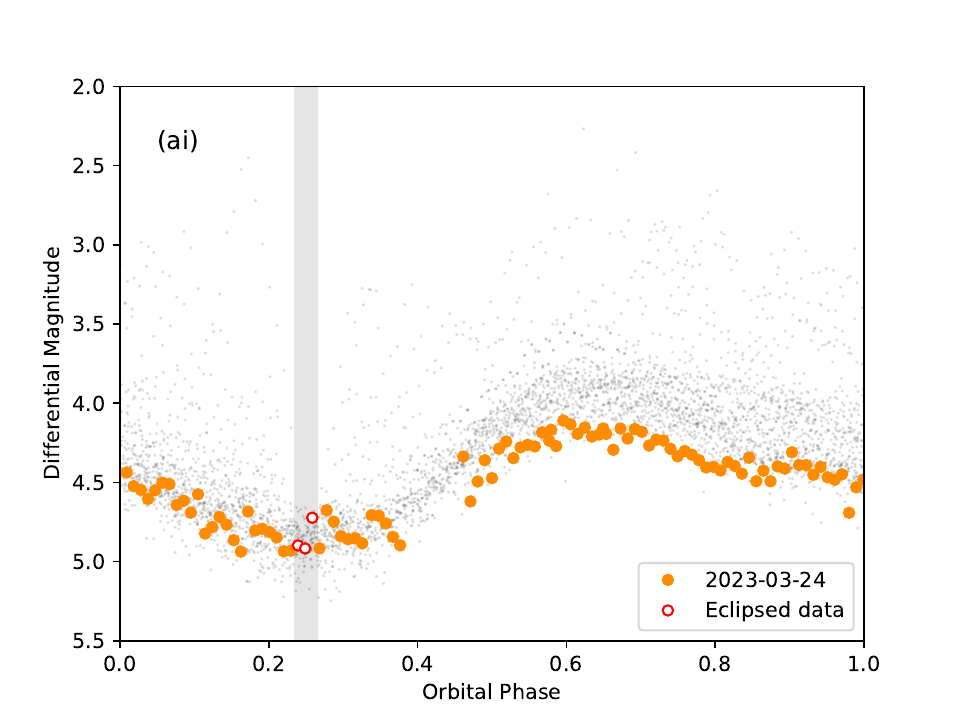}
	\end{subfigure}
	\begin{subfigure}{}
	\includegraphics[width=0.44\textwidth]{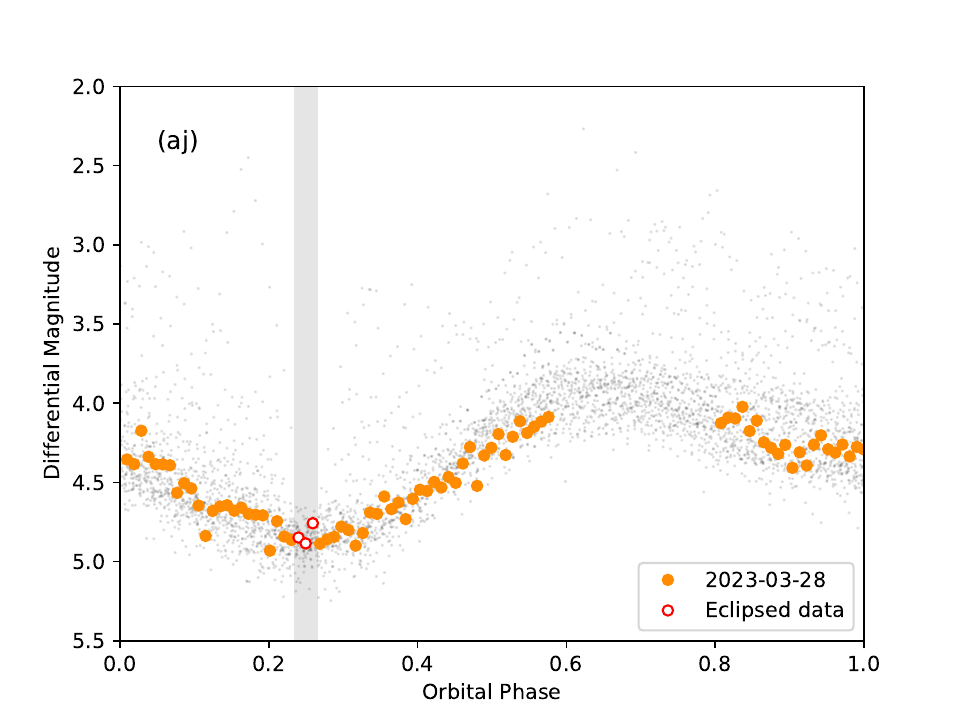}
	\end{subfigure}
\centering
\caption{\textit{continued}}
\end{figure}

\setcounter{figure}{6}

\begin{figure}
	\begin{subfigure}{}
	\includegraphics[width=0.44\textwidth]{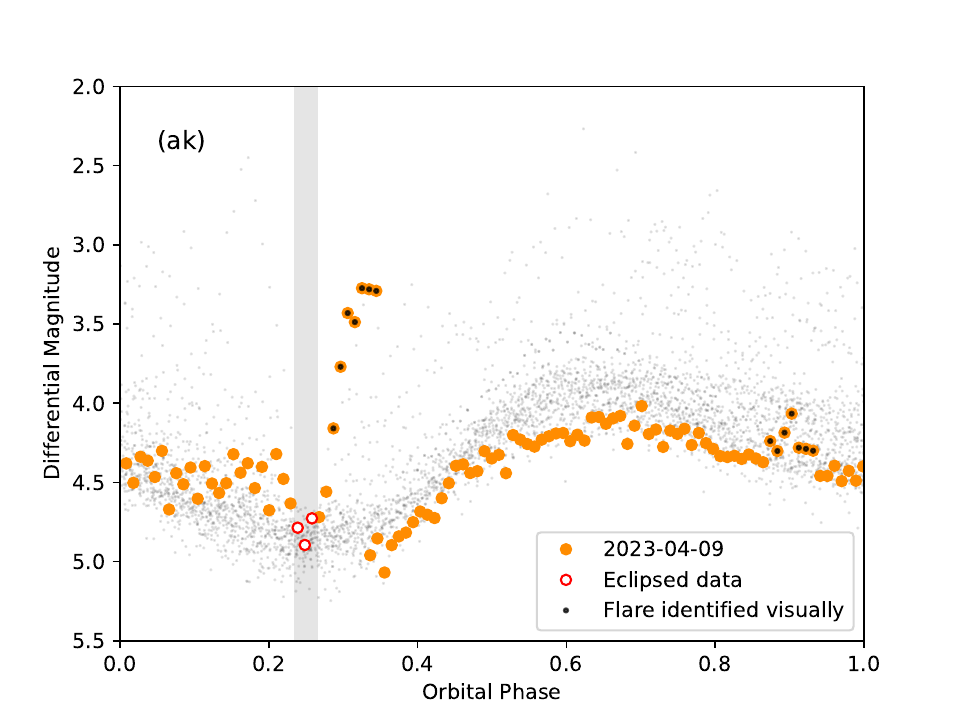}
	\end{subfigure}
	\begin{subfigure}{}
	\includegraphics[width=0.44\textwidth]{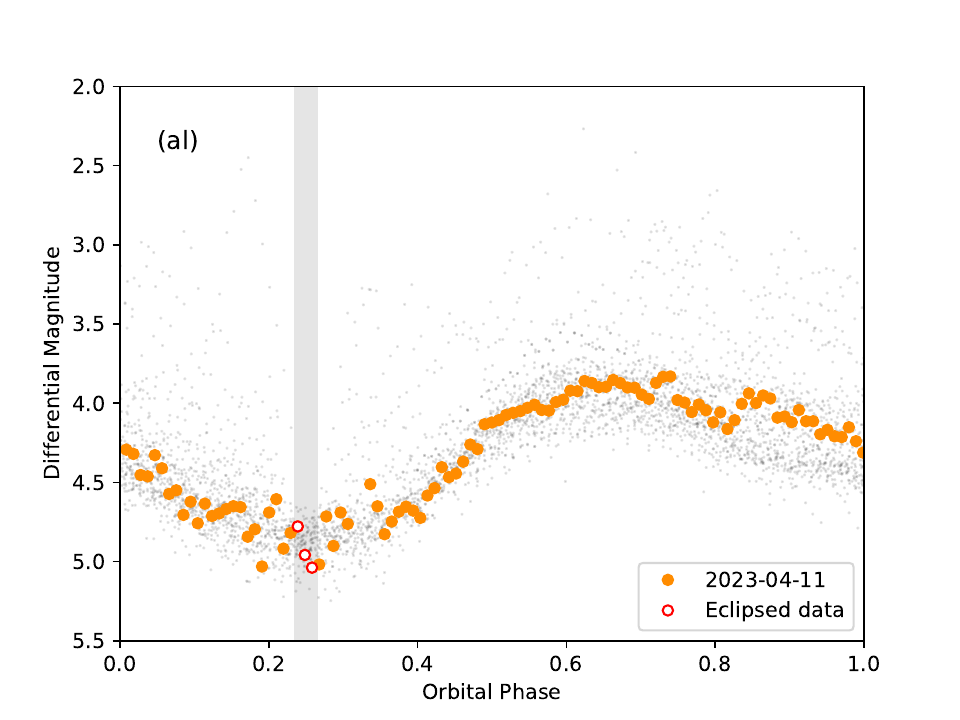}
	\end{subfigure}
	\begin{subfigure}{}
	\includegraphics[width=0.44\textwidth]{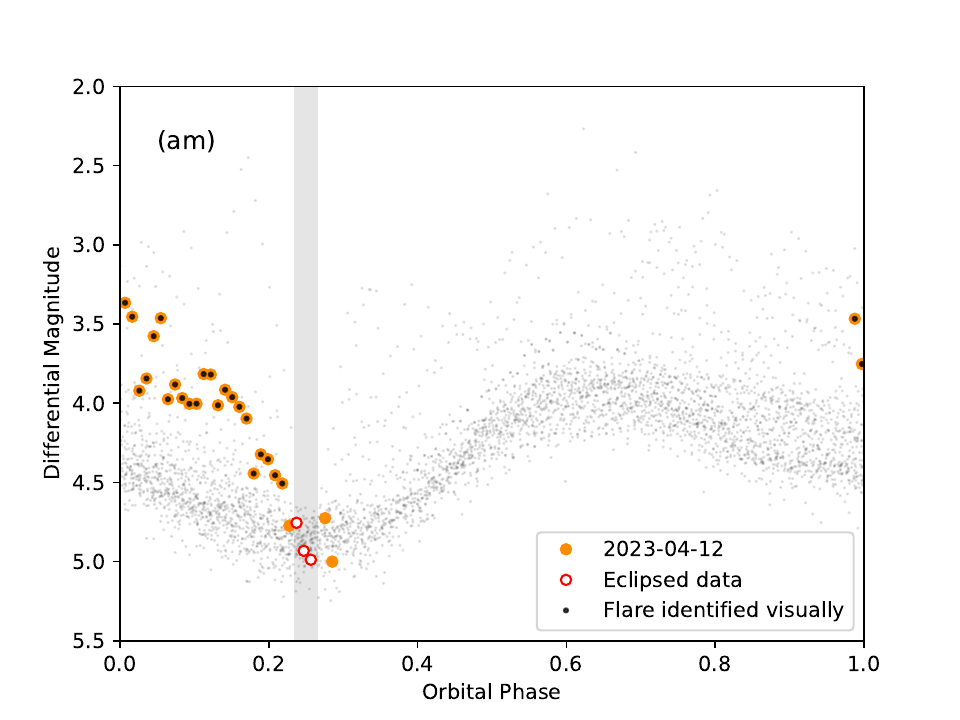}
	\end{subfigure}
	\begin{subfigure}{}
	\includegraphics[width=0.44\textwidth]{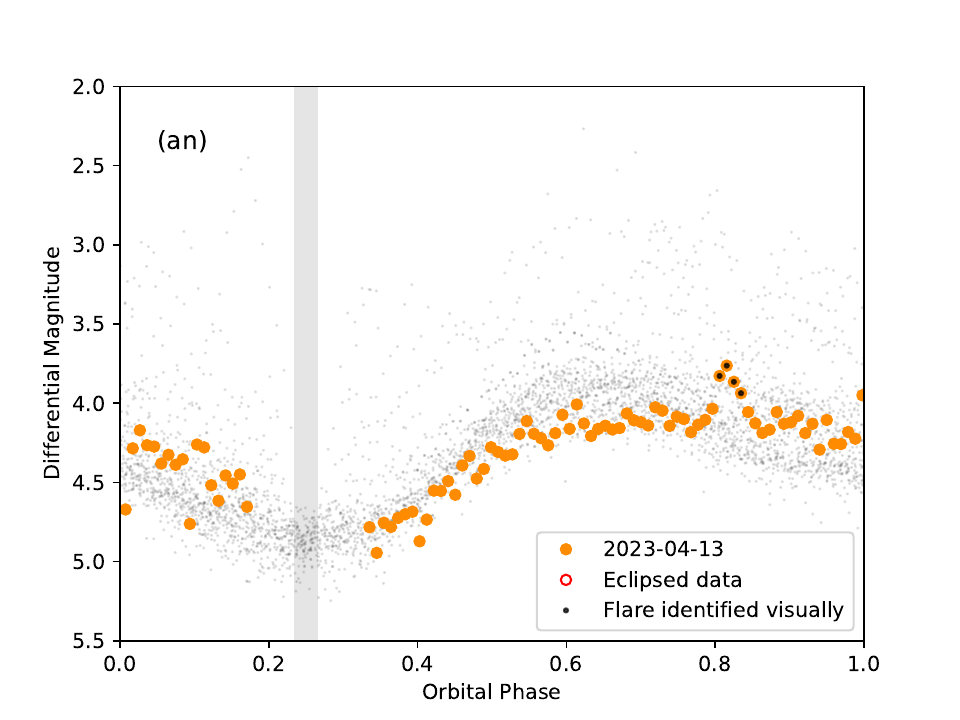}
	\end{subfigure}
	\begin{subfigure}{}
	\includegraphics[width=0.44\textwidth]{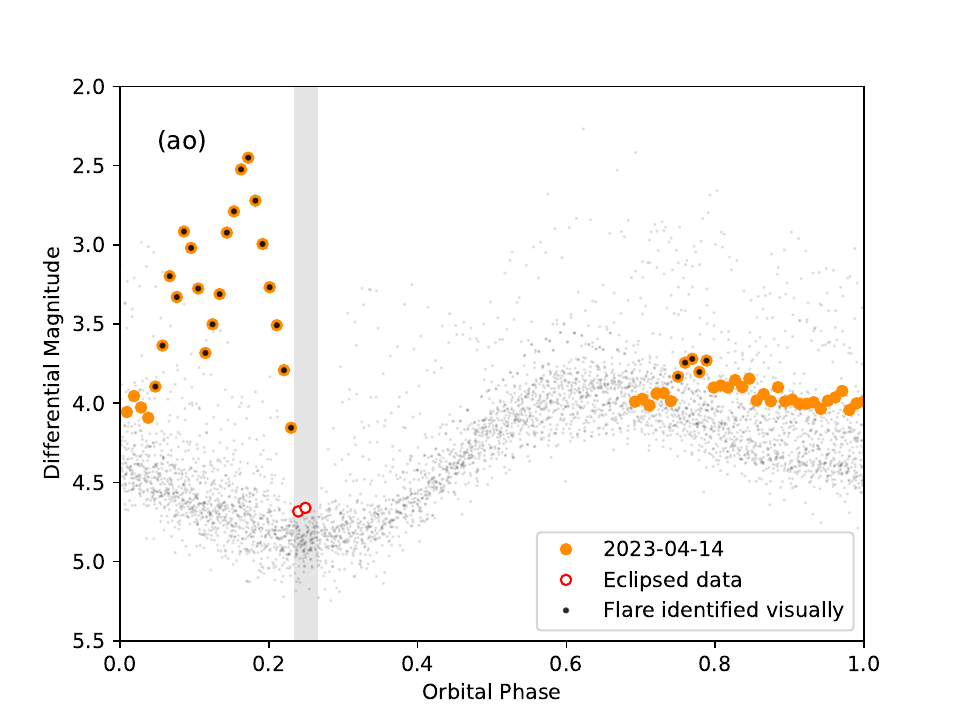}
	\end{subfigure}
	\begin{subfigure}{}
	\includegraphics[width=0.44\textwidth]{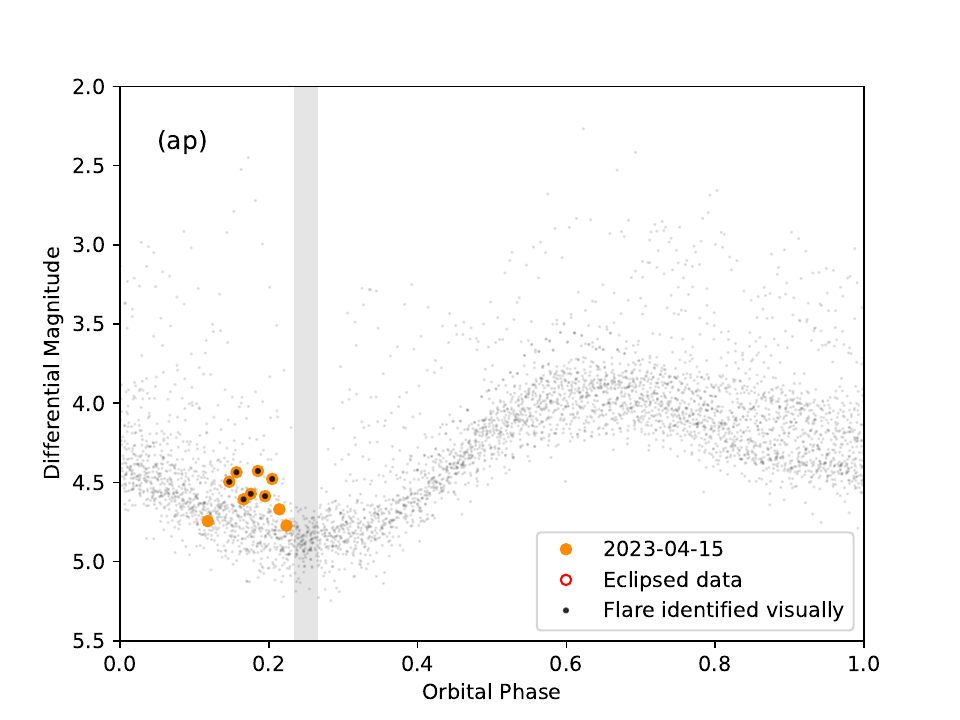}
	\end{subfigure}
\centering
\caption{\textit{continued}}
\end{figure}

\setcounter{figure}{6}

\begin{figure}
	\begin{subfigure}{}
	\includegraphics[width=0.44\textwidth]{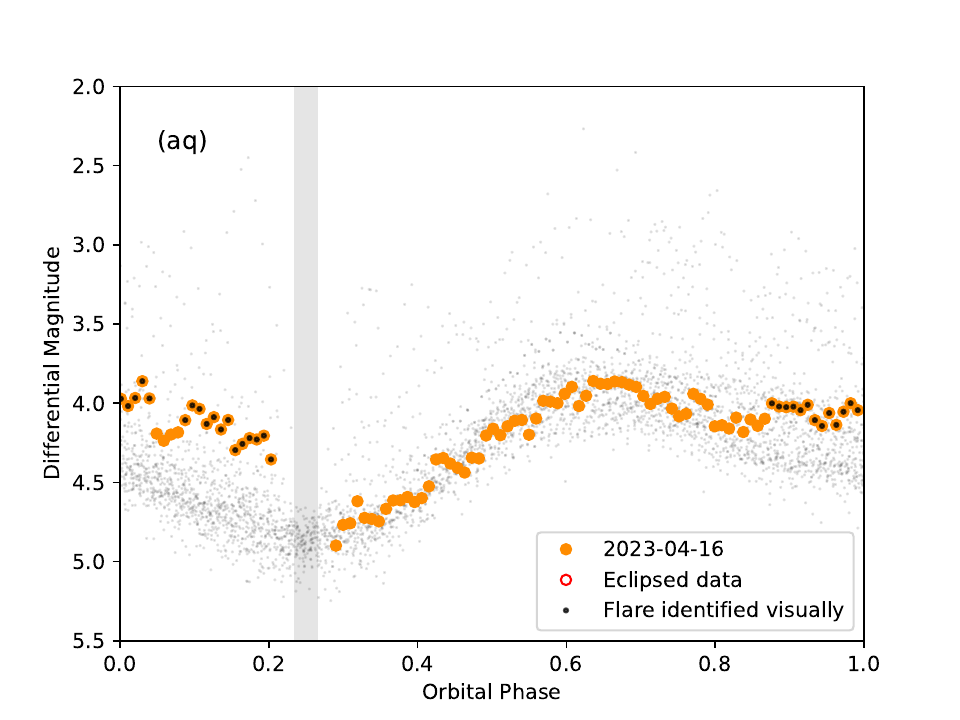}
	\end{subfigure}
	\begin{subfigure}{}
	\includegraphics[width=0.44\textwidth]{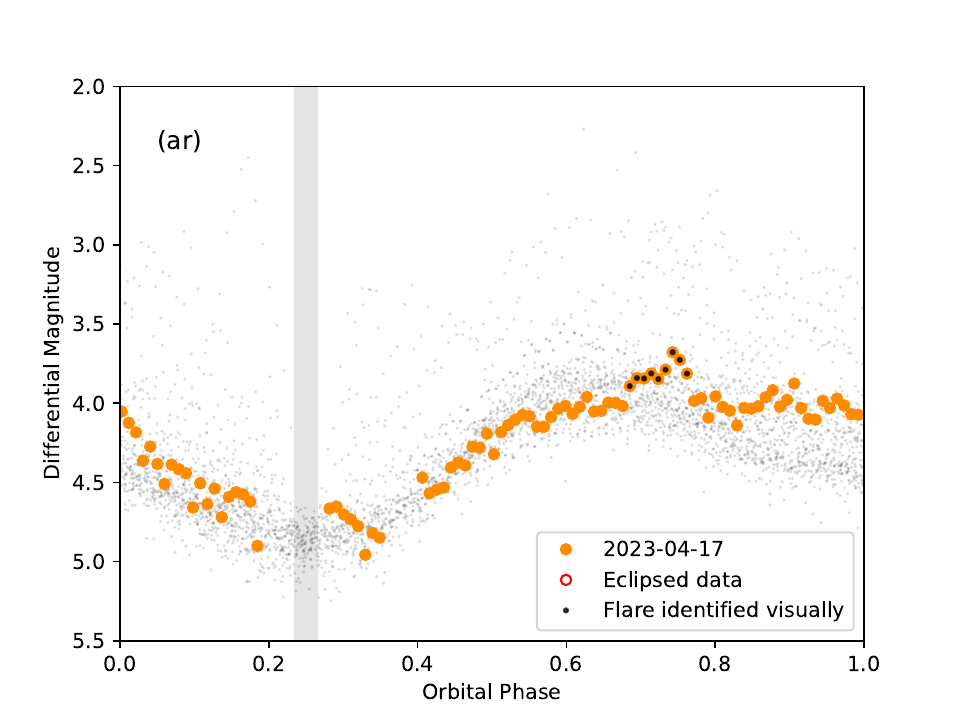}
	\end{subfigure}
	\begin{subfigure}{}
	\includegraphics[width=0.44\textwidth]{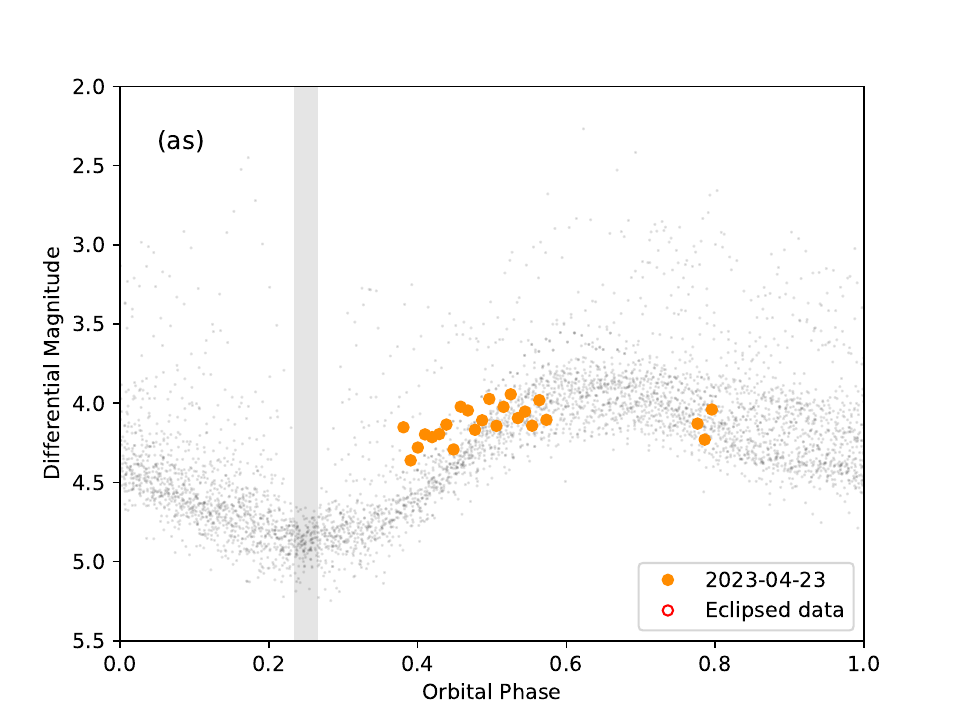}
	\end{subfigure}
	\begin{subfigure}{}
	\includegraphics[width=0.44\textwidth]{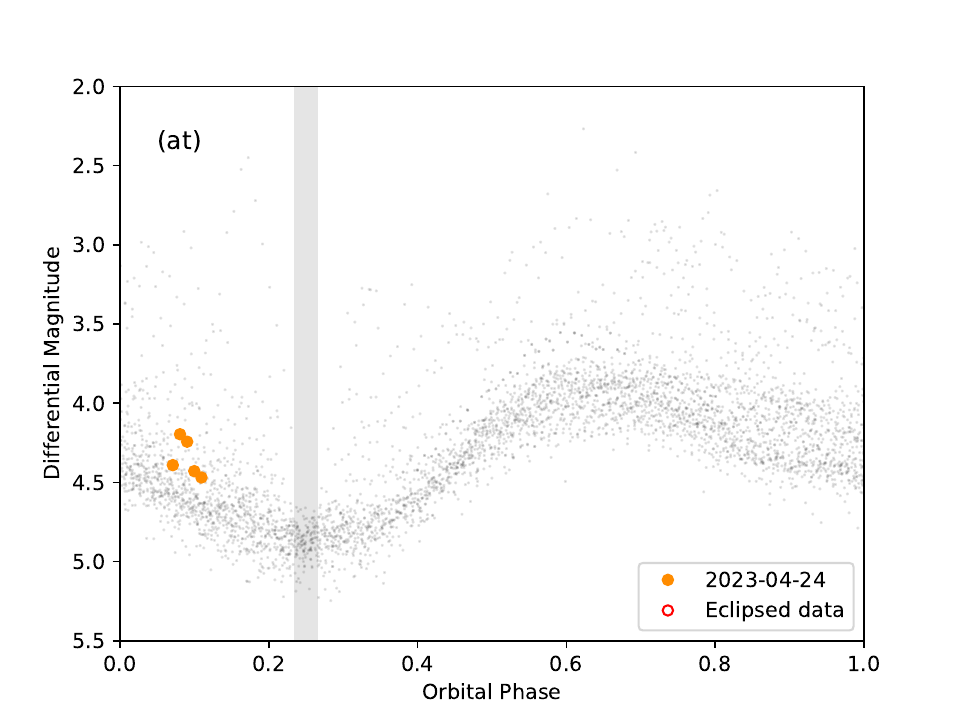}
	\end{subfigure}
	\begin{subfigure}{}
	\includegraphics[width=0.44\textwidth]{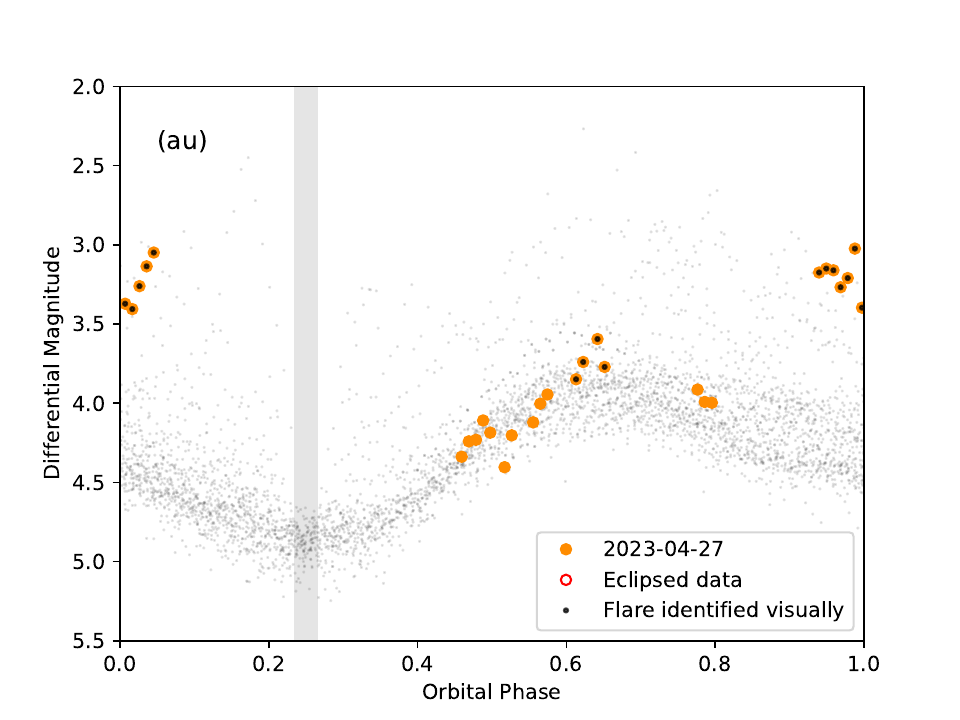}
	\end{subfigure}
	\begin{subfigure}{}
	\includegraphics[width=0.44\textwidth]{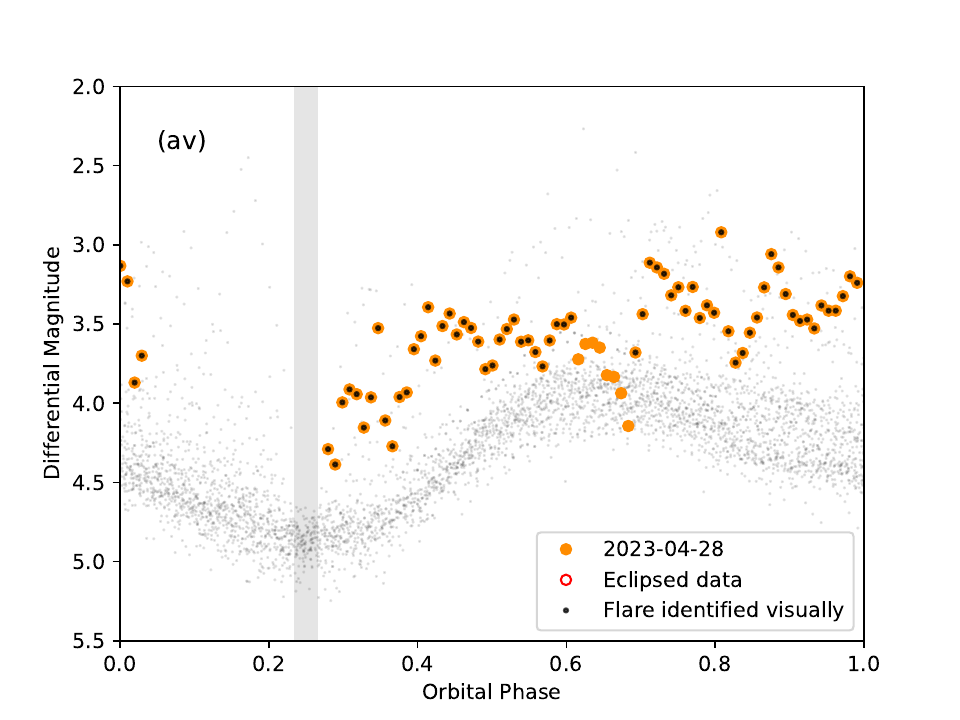}
	\end{subfigure}
\centering
\caption{\textit{continued}}
\end{figure}

\clearpage
\section{$\bm{K}$-Correction for the Radial Velocity Curve}
\label{app:kcorr}
A simulation method was applied to calculate the $K$-correction effect on the radial velocity (RV) curve of \src\ based on an algorithm similar to the ELC code \citep{2000A&A...364..265O}. In the simulation, a companion star in a binary system of a mass ratio $Q$ is filling the Roche lobe with a factor $f$ centred at the origin of a Cartesian coordinate system (see the definitions of $f$ and $Q$ in the main text), and a compact star (i.e., a neutron star in the case of \src) is located at ($a$, 0, 0). We made a mesh of 10,000 points randomly generated with a spherically symmetric distribution on the companion stellar surface ($P_i\,\hat{p}_i$ for the $i$-th point) and defined: (i) the normalized normal vector to the corresponding stellar surface as $\hat{n}_i$; (ii) the position vector from the surface to the neutron star as $D_i\,\hat{d}_i$; and (iii) the normalized line-of-sight vector to Earth as $\hat{l}$. The generated points become grids for the companion emission computation in the following procedure. For a perfect spherical surface, the areas of the grids, $s_i$, are statistically the same, given the spherically symmetric distribution. For non-spherical surfaces, the grid areas can be approximated by
\begin{equation}
s_i \propto \frac{P_i^2}{\hat{n}_i \cdot \hat{p}_i}.
\end{equation}

The surface temperature of the $i$-th grid, $T_i$, was computed based on both the intrinsic stellar temperature, $T_{\mathrm{in},i}$, and the irradiation from the pulsar. The intrinsic temperature on the surface varies according to the local surface gravity of the star, $g_i$, known as gravity darkening, which can be modelled by the von Zeipel's theorem \citep{1924MNRAS..84..665V,1999A&A...347..185M},
\begin{equation}
T_{\mathrm{in},i} = T_\mathrm{in,pole}\left(\frac{g_i}{g_\mathrm{pole}}\right)^{\beta},
\end{equation}
where $T_\mathrm{in,pole}$ and $g_\mathrm{pole}$ are the intrinsic temperature and the surface gravity at the pole, i.e., on the surface of (0, 0, $z_\mathrm{pole}$), and $\beta=0.08$ for low-mass stars with convective envelopes \citep{1967ZA.....65...89L}. The polar temperature and the effective temperature of the star, $T_\mathrm{in,eff}$, can be related by the Stefan–Boltzmann law and thus can be written as \citep{2000A&A...364..265O}
\begin{equation}
\begin{aligned}
T_\mathrm{in,pole}^4 &= T_\mathrm{in,eff}^4\left(\frac{\oint ds}{\oint (g_i/g_\mathrm{pole})^{4\beta} ds}\right)\\
&\approx T_\mathrm{in,eff}^4\left(\frac{\sum{s_i}}{\sum{(g_i/g_\mathrm{pole})^{4\beta} s_i}}\right),
\end{aligned}
\end{equation}
where $ds$ is the differential element of the stellar surface.

The temperatures of the irradiated surfaces were computed based on the $R$-function method \citep{1990ApJ...356..613W}. In the original version of the algorithm, the $R$-functions are defined as
\begin{equation}
\begin{aligned}
R_{1,i} = 1+\frac{F^\prime_{2,i}}{F_{1,i}}\\
R_{2,i} = 1+\frac{F^\prime_{1,i}}{F_{2,i}}
\end{aligned},
\end{equation}
where $F_{1,i}$ (or $F_{2,i}$) is the intrinsic flux of the surface element $i$ of the primary (secondary) star without irradiation, and $F^\prime_{2,i}$ (or $F^\prime_{1,i}$) is the irradiation flux to the surface element from the secondary (primary) star. As $F^\prime_{2,i}/F_{1,i}$ is also a function of $R_{2,i}$ (and so are $F^\prime_{1,i}/F_{2,i}$ and $R_{1,i}$), an iteration scheme was proposed by \cite{1990ApJ...356..613W} to obtain the $R$-functions. In \src, we can simply ignore the irradiation effect to the neutron star from the companion (i.e., $R_{1,i}=1$) and get $R_{2,i}$ ($R_i$ hereafter) without iteration using
\begin{equation}
\begin{aligned}
F^\prime_{1,i} &= \frac{\alpha L_\mathrm{irr}(\hat{n}_i \cdot \hat{d}_i)}{4\pi D_i^2}\\
F_{2,i} &= (1-\frac{x}{3})\,\sigma T_{\mathrm{in},i}^4
\end{aligned},
\end{equation}
where $\alpha=0.5$ is the albedo for low-mass stars with convective envelopes \citep{2001MNRAS.327..989C}, $L_\mathrm{irr}$ is the irradiation power from the pulsar, $x$ is the bolometric linear limb-darkening coefficient \citep{1993AJ....106.2096V}, and $\sigma$ is the Stefan–Boltzmann constant. The $R$-factor can then be used to find the final temperatures of the heated surfaces by
\begin{equation}
T_i = T_{\mathrm{in},i}\,R_i^\frac{1}{4}.
\end{equation}

\begin{figure}
\includegraphics[width=0.9\textwidth]{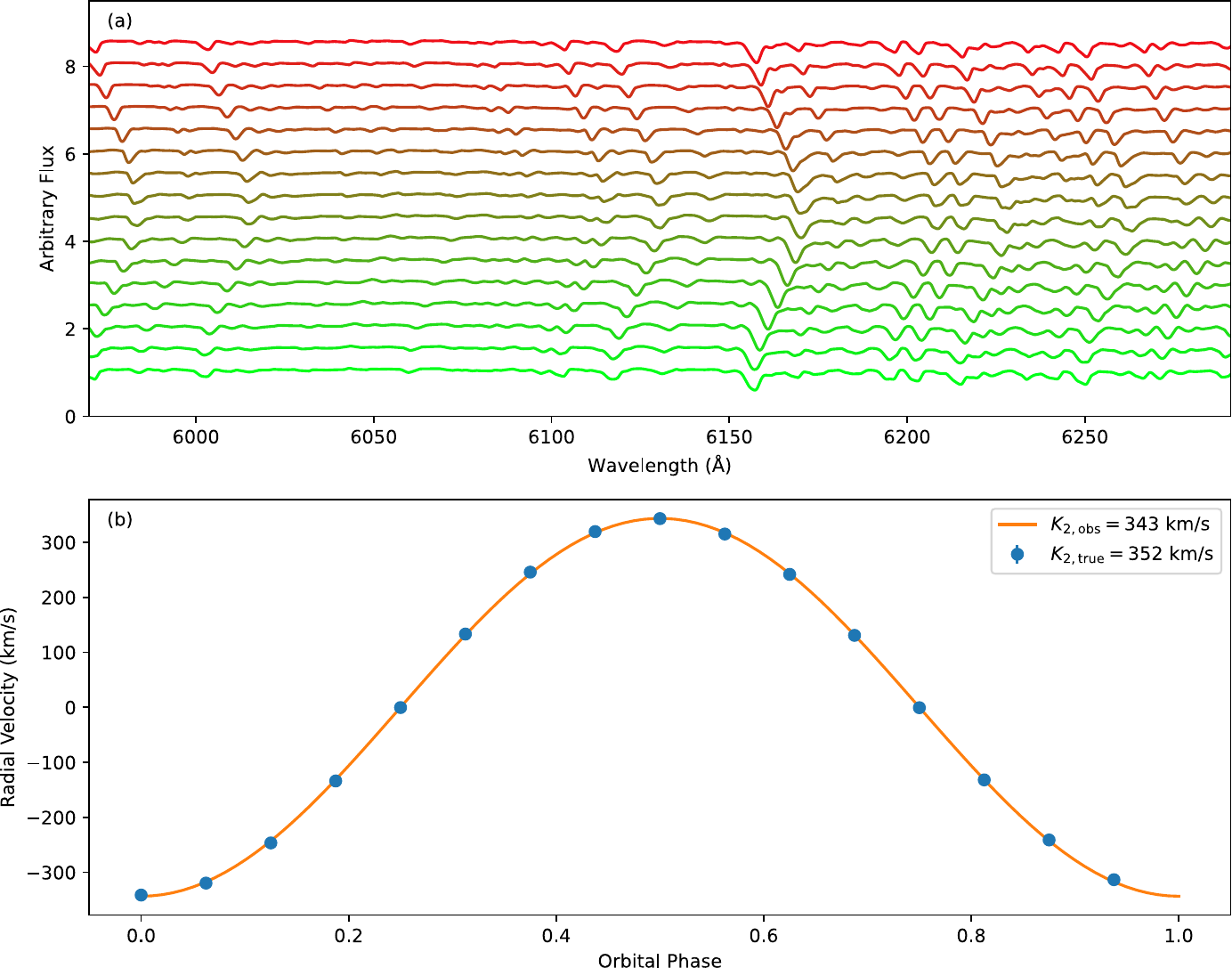}
\centering
\caption{Example of the radial velocity curve simulation of \src. The upper panel shows the 16 simulated spectra at different orbital phases (the bottom one refers to phase zero), with an assumption of $K_\mathrm{2,true}=352$\kms. The lower panel is the ``observed'' radial velocity curve with $K_\mathrm{2,real}=343$\kms, which is consistent with the VLT spectroscopic observations.}
\label{fig:rv_sim}
\end{figure}

With the information computed above, we theoretically synthesized the observed spectrum of the irradiated companion at different orbital phases. We employed the high-resolution synthetic specific intensity spectra with solar abundances by the PHOENIX atmosphere code \citep{2013A&A...553A...6H} downloaded from the G{\"o}ttingen Spectral Library\footnote{\url{https://phoenix.astro.physik.uni-goettingen.de}} as the building blocks. The temperature and surface gravity ranges of the obtained PHOENIX synthetic spectra are 3100--12000~K (with a step of 200~K, except for the step from 6900~K to 7000~K) and 2.0--6.0 (a step of 0.5; in cgs units), which cover the surface temperatures and local gravities of the companion's grid elements for \src. For each combination of temperature and gravity, there are 78 synthetic spectra viewed at different angles (written as $\mu_i = \cos \Theta_i = \hat{n}_i \cdot \hat{l}$ here, where $\Theta_i$ is the angle between the normal vector of the grid and the line of sight of the observer) for the consideration of the limb-darkening effect. At a given orbital phase, linear interpolation was employed on the PHOENIX data to obtain the spectra of the companion grids at the computed temperatures ($T_i$), gravities ($g_i$), and viewing angles ($\mu_i$). Lastly, Doppler shifting due to the orbital motion and rotation of the companion was applied on the grids' spectra, which were then weighted by $s_i\mu_i$ (i.e., the projected areas of the grids) and summed together to yield the final simulated spectrum at the phase. 

To find the $K$-correction of \src\ (i.e., $K_{2,\mathrm{obs}}$ and $K_{2,\mathrm{true}}$ are the observed/true RV semi-amplitude of the companion, respectively) during the \textit{Very Large Telescope} (VLT) spectroscopic observations \citep{2021A&A...649A.120M}, we simulated spectra at 16 different orbital phases uniformly distributed along the orbit. The spectral ranges employed are 5970--6291\AA\ and 6421--6522\AA, which are the same wavebands used in the VLT analysis \citep{2021A&A...649A.120M}. We then fitted a sinusoidal function to the obtained RV curve to get $K_{2,\mathrm{obs}}$. It is worth noting that the RV curves slightly deviate from sinusoidal profiles because of the non-spherical shapes of the companion as well as the uneven heating on the companion, and these small deviations introduce systematic errors to $K_{2,\mathrm{obs}}$. 

For the binary model constructions in the RV simulations, we assumed that (i) the binary geometry of \src\ remained the same in the epochs of the LOT and VLT observations \citep{2018ApJ...866...71C,2019A&A...621L...9Y} and (ii) the irradiation power of the pulsar ($L_\mathrm{irr}$) resulted in a peak-to-peak amplitude of 0.8--1.0~mag in the $R$ band during the VLT observation (the amplitude was reported as $\sim$0.9~mag; \citealt{2018ApJ...866...71C}). The first assumption allows us to infer the binary geometries of \src\ with the LOT dataset (with very low irradiation) and a range of assumed $K_{2,\mathrm{true}}$ using ELC. The second assumption gives us a range of $L_\mathrm{irr}$ (corresponding to the $R$-band amplitude range from 0.8 to 1.0~mag) to model the irradiation. Searching from $K_{2,\mathrm{true}}=344$ to $364$\kms, we find $K_{2,\mathrm{true}}=353.5\pm6.5$\kms, which results in observed values consistent with the VLT measure (i.e., $K_2=343.3\pm4.4$\kms; Figure \ref{fig:rv_sim}).
This $K_{2,\mathrm{true}}$ is also consistent with the result from the ``empirical $K$-correction'' for \src\ using GTC spectroscopy (i.e., $K_2=340-375$\kms; \citealt{2025MNRAS.536.2169S}), supporting the validity.

We notice that this result disagrees with the claim by \cite{2021A&A...649A.120M} that $K_{2,\mathrm{true}}<343.3$\kms. This is probably because the quenching of the absorption lines on the day side of the companion \citep{1988ApJ...324..411W} is insignificant for \src\ in the spectral ranges considered (Figure \ref{fig:rv_sim}).

\section{\textit{Chandra} Observations}
\label{app:chandra}
\begin{figure}
\includegraphics[width=0.9\textwidth]{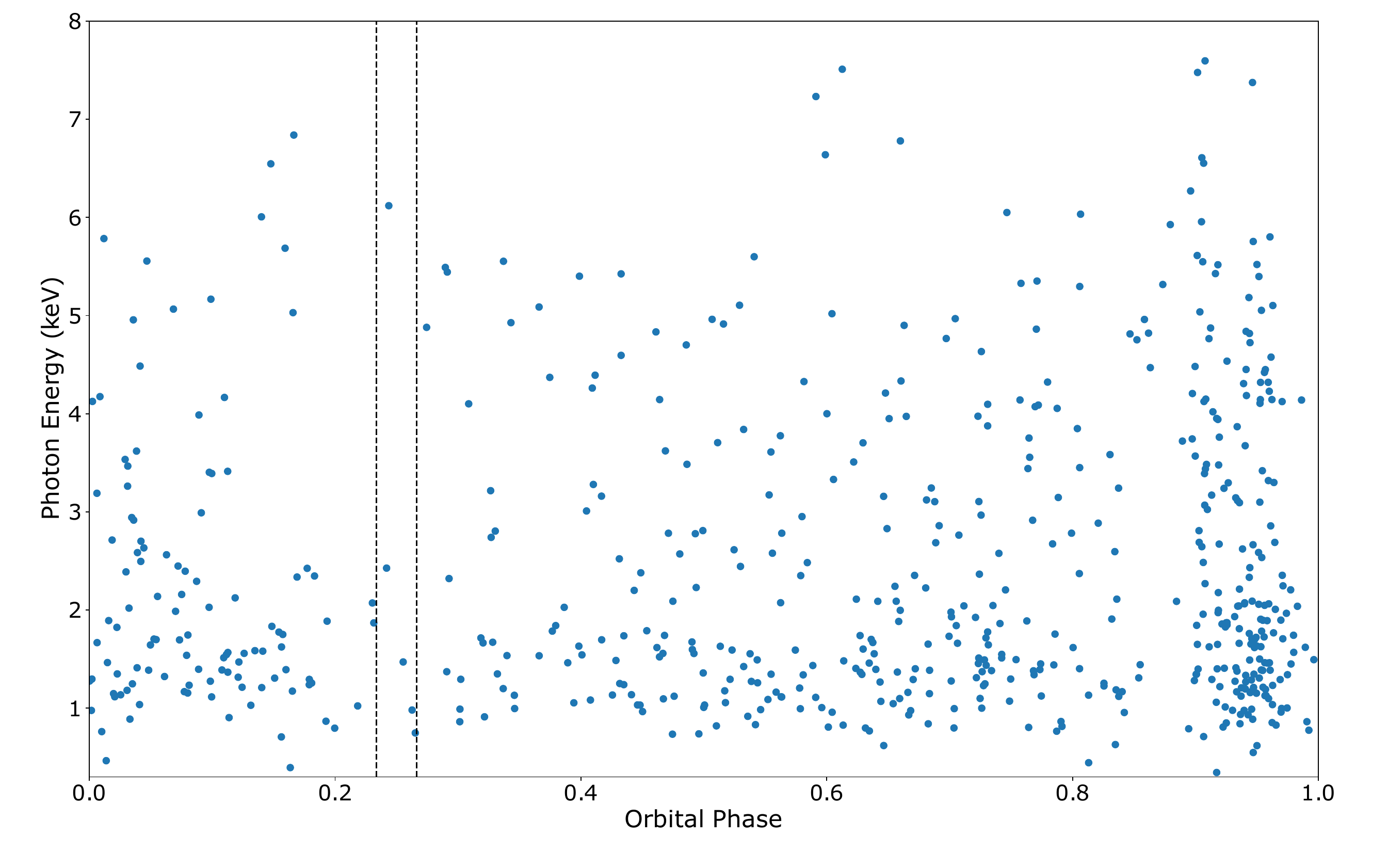}
\centering
\caption{The X-ray energy v.s. orbital phase of the \textit{Chandra}-detected photons of \src. The dashed lines indicate the eclipse region, in which five photons were detected. According to the X-ray image analysis, there are about six background counts in total.}
\label{fig:chandra_lc}
\end{figure}

Using three \textit{Chandra} archival observations of \src \bb{\dataset[(DOI: 10.25574/cdc.657)]{https://doi.org/10.25574/cdc.657}} with exposure times of 23.1 ks (ObsID 19037), 25.1 ks (ObsID 19038), and 22.9 ks (ObsID 19039), we checked if \src\ showed eclipses in X-rays. The observations were reduced and analysed with \texttt{CIAO} (version 4.13).  The task \texttt{dmcopy} was employed to extract source photons from the three observations with circular regions of radii 3$\farcs$0, 2$\farcs$5, and 2$\farcs$3, respectively, based on visual inspection on the X-ray images. The time columns of the source events were then barycentric corrected by \texttt{axbary}. Figure \ref{fig:chandra_lc} shows the source events in a plane of photon energy against orbital phase with the flare eclipse region indicated by two dashed lines. Five photons were detected during the flare eclipses, and three of them have energies higher than 1~keV. According to the source-free regions of the X-ray images, the expected number of the background counts is six in total. These strongly suggest that the five photons are not from the soft X-ray background but \src, and we therefore conclude that there was no X-ray eclipse observed in the \textit{Chandra} data.

\bibliography{j1048_rift_nat.bib}
\end{document}